\documentclass[twocolumn]{aastex631}

\newcommand\aastex{AAS\TeX}

\defcitealias{1991MNRAS.251..545H}{H91}
\usepackage{comment}
\usepackage[normalem]{ulem}  % for strikeout, for referee

\received{February 28, 2024}
\revised{December 16, 2024}
\accepted{December 16, 2024}

\shorttitle{On the Red Clump Age Method}
\shortauthors{Chriss and Worthey}
\begin{document}

\title{On the Age Calibration of Open Clusters using Red Clump Stars}

\author[0009-0003-8929-7110]{Abigail R. Chriss}
\affiliation{Department of Physics and Astronomy, Bowdoin College, 8800 College Station, Brunswick, ME 04011-8488, USA}

\author[0000-0003-1388-5525]{Guy Worthey}
\affiliation{Department of Physics and Astronomy, Washington State University, 1245 Webster Hall, Pullman, WA 99164-2814, USA}

%% Note that the \and command from previous versions of AASTeX is now
%% depreciated in this version as it is no longer necessary. AASTeX 
%% automatically takes care of all commas and "and"s between authors names.

%% AASTeX 6.31 has the new \collaboration and \nocollaboration commands to
%% provide the collaboration status of a group of authors. These commands 
%% can be used either before or after the list of corresponding authors. The
%% argument for \collaboration is the collaboration identifier. Authors are
%% encouraged to surround collaboration identifiers with ()s. The 
%% \nocollaboration command takes no argument and exists to indicate that
%% the nearby authors are not part of surrounding collaborations.

%% Mark off the abstract in the ``abstract'' environment. 
\begin{abstract}

In this study, we extend the dust-independent \citet{1991MNRAS.251..545H} relation between cluster age and $d_{B-R}$ color difference between the red giant branch (RGB) and red clump to younger cluster ages. We perform membership analysis on fourteen galactic open clusters using \textit{Gaia} DR3 astrometry, then compute the difference in color of the RGB and red clump $d_{B-R}$ using \textit{Gaia} photometry. We also compute $d_{B-R}$ for five fields surrounding Small Magellanic Cloud (SMC) clusters. We find that the trend derived from older clusters does not extrapolate to younger ages and becomes double-valued. We confirm that $d_{B-R}$ is independent of metallicity. Current stellar evolutionary isochrones do not quantitatively reproduce the trend and furthermore predict an increased color gap with a decrease in metallicity that is not echoed in the data. Integrated light models based on current isochrones exaggerate the color change over the $-0.5 <$ [Fe/H] $< 0$ interval at the few-percent level.

\end{abstract}

%% Keywords should appear after the \end{abstract} command. 
%% The AAS Journals now uses Unified Astronomy Thesaurus concepts:
%% https://astrothesaurus.org
%% You will be asked to selected these concepts during the submission process
%% but this old "keyword" functionality is maintained in case authors want
%% to include these concepts in their preprints.

%% From the front matter, we move on to the body of the paper.
%% Sections are demarcated by \section and \subsection, respectively.
%% Observe the use of the LaTeX \label
%% command after the \subsection to give a symbolic KEY to the
%% subsection for cross-referencing in a \ref command.
%% You can use LaTeX's \ref and \label commands to keep track of
%% cross-references to sections, equations, tables, and figures.
%% That way, if you change the order of any elements, LaTeX will
%% automatically renumber them.
%%
%% We recommend that authors also use the natbib \citep
%% and \citet commands to identify citations.  The citations are
%% tied to the reference list via symbolic KEYs. The KEY corresponds
%% to the KEY in the \bibitem in the reference list below. 

\section{Introduction} \label{section_intro}

Star clusters represent both a crucial testing ground for the theory of stellar evolution and markers for the chemical and dynamical history of the Milky Way and other galaxies. While massive star clusters typically host multiple stellar populations \citep{2012A&ARv..20...50G}, lower-mass open clusters are currently presumed to be simple stellar populations (SSPs) of a single age and heavy element abundance pattern. Cluster ages are often derived from the color magnitude diagram (CMD), where, especially, the luminosity of the main sequence turnoff (MSTO) provides a theoretically robust chronometer \citep{1995ApJ...444L...9C}.

Even assuming perfect comparison models, distance uncertainty and line of sight dust extinction add error to age estimates. 
\citet{1991MNRAS.251..545H}, hereafter \citetalias{1991MNRAS.251..545H}, proposed a method that bypasses both dust and distance. The color of the He-burning red clump is subtracted from the color of the H-burning red giant branch (RGB) at the same luminosity. \citetalias{1991MNRAS.251..545H} used Johnson-Cousins $B$ and $R$ filters and thus the color difference is $d_{B-R}$. Age was found to track $d_{B-R}$ for clusters older than $\sim$2 Gyr and for a variety of heavy element abundances. A bonus advantage of this method is that red clump stars are much brighter than MSTO stars, and thus the method could be applied to distant clusters.

The advent of star formation history reconstruction methods, where the entire CMD is fit with a swarm of stellar evolutionary isochrones \citep{2001ApJS..136...25H,2002MNRAS.332...91D}, might explain why the \citetalias{1991MNRAS.251..545H} formula has not seen wide use \citep{2016ARA&A..54...95G}. Girardi notes, however, that the reconstructions for older ages rest upon red clump and RGB lifetimes, something predicted by stellar evolutionary theory only at the 20\% level.

The purpose of the present work is to confirm and extend \citetalias{1991MNRAS.251..545H}'s result. A red clump should exist in the CMD for ages as young as $\sim$150 Myr, corresponding to MSTO masses of $\sim$5 M$_\odot$. We were curious if \citetalias{1991MNRAS.251..545H}'s result extended to younger populations. To that end, we mined \textit{Gaia} open cluster data for additional clusters with red clumps, with particular attention to clusters between 500 Myr and 2 Gyr in age.

We describe cluster membership discrimination and our derivation of isochrone-based ages in $\S$\ref{section_methods}. The \citetalias{1991MNRAS.251..545H} method is addressed in $\S$\ref{section_results}, and we discuss the result in $\S$\ref{section_discussion}.

\section{Cluster Sample and Membership Analysis} \label{section_methods}

From open cluster lists \citep{2013A&A...558A..53K,2020A&A...640A...1C} we selected clusters beyond the \citetalias{1991MNRAS.251..545H} list, initially in the age range 200 Myr to 2 Gyr. As the project developed, we added a few older clusters to test the repeatability of the \citetalias{1991MNRAS.251..545H} method. Ideal clusters lie nearby and bloom richly with stars.

\begin{figure}
	\includegraphics[width=\columnwidth]{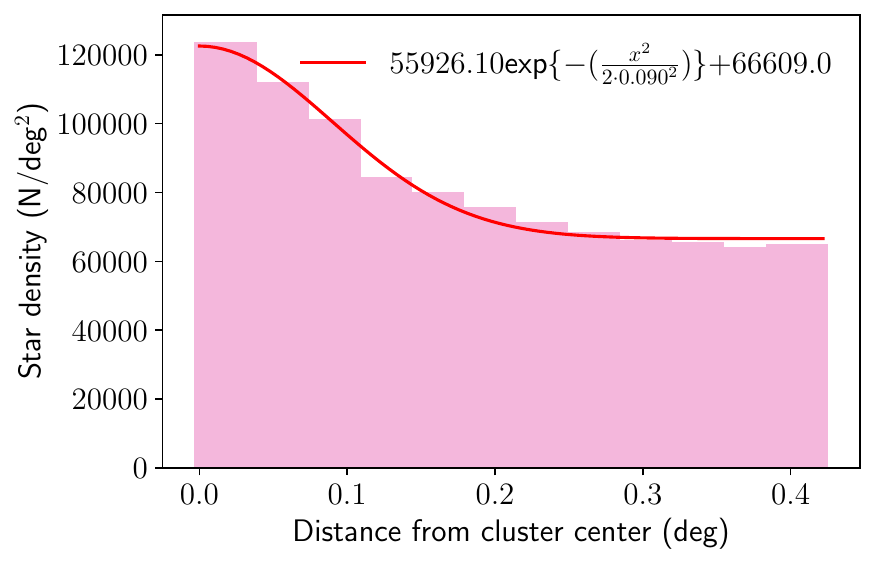}
    \caption{Stellar density in annuli proceeding from the center of the sky position of NGC~7789. A Gaussian model is shown (red). Field star density is taken as the constant-density asymptote.}
    \label{fig:pos_dist}
\end{figure}

For each of the fourteen galactic open clusters presented in Table \ref{tab:cluster BP-RP new} we performed a membership analysis before applying reddening corrections and analyzing the CMD. The probabilities of membership for stars in each cluster were estimated using position ($\alpha , \delta$), proper motion ($\mu_\alpha , \mu_\delta$), and parallax ($\pi$) from \textit{Gaia} Data Release 3 (DR3). Cluster centers, proper motion central locations, and parallaxes from SIMBAD served as a first guess for each cluster, though we allowed drifts from these values during the fitting process. For position and proper motion, we drew annuli around the central location and computed star density (number per square degree or number per km s$^{-1}$, respectively) as a function of annulus bin radius ($r$). We employed a least squares fitter to model functions for the total stars and the field star background. We modeled the cluster spatial profile as a Gaussian with a constant-density pedestal representing field stars (Figure \ref{fig:pos_dist}). Based on position, for example, the cluster membership probability for each star $i$ is 

$$
P_{{\rm cluster,\ }(\alpha, \delta)}(i) = \frac{N_{\rm total}(r_i) - N_{\rm field}(r_i)}{N_{\rm total}(r_i)},
$$ where $r_i$ is the distance of a star from the cluster center, and the number of counts per square degree $N$ come from two (cluster and field) function fits to the density profile. In other words, the probability is the expected fraction of member stars as a function of distance from the cluster center.

\begin{table*}
\centering
\caption{Age (log years) and errors from \citet{2020A&A...640A...1C}, color excess (magnitude) from \citet{2013A&A...558A..53K}, and color difference of red clump and RGB and errors (magnitude) for Milky Way open clusters in this study. Number of red clump stars with probability greater than 0.6 and opening angle for cone search (degrees) are also included. Metallicity from \citet{2021MNRAS.504..356D} for Ruprecht 68 and NGC 2509, and from \citet{2013A&A...558A..53K} for all other clusters.}
\label{tab:cluster BP-RP new}
\begin{tabular}{lcccccccc}
\hline
Name & Age  & E($BP-RP$) & [Fe/H] & $d_{BP-RP}$  & $d_{B-R}$ & N$_{\text{RC}}$(60) & Opening angle \\
& (log years) & (mag) & (dex) & (mag) & (mag) & & (deg) & \\
\hline
    Melotte 71 & 8.99 $\pm \ 0.175$ & 0.139 & -0.22 & 0.118 $\pm \ 0.035 $ & 0.208 $\pm \ 0.046$ & 8  & 0.224\\
    King 5    & 9.01 $\pm \ 0.175$ & 0.897 & -0.30 & 0.104 $\pm \ 0.012$ & 0.185 $\pm \ 0.016$ & 8 & 0.272\\
    NGC 2477  & 9.05 $\pm \ 0.175$ & 0.390 & -0.192 & 0.176 $\pm \ 0.009$ & 0.311 $\pm \ 0.013$ & 58 & 0.450\\
    NGC 1245  & 9.08 $\pm \ 0.175$ & 0.335 & 0.10 & 0.107 $\pm \ 0.009$ & 0.189 $\pm \ 0.012$ & 38 & 0.326\\
    NGC 6208  & 9.15 $\pm \ 0.175$ & 0.279 & -0.03 & 0.082 $\pm \ 0.010$ & 0.146 $\pm \ 0.014$ & 6 & 0.486\\
    NGC 2509  & 9.18 $\pm \ 0.175$ & 0.139 & 0.082 & 0.098 $\pm \ 0.015$ & 0.175 $\pm \ 0.020$ & 3 & 0.179\\
    NGC 7789  & 9.19 $\pm \ 0.175$ & 0.296 & -0.24 & 0.113 $\pm \ 0.006$ & 0.195 $\pm \ 0.010$ & 35 & 0.422\\
    NGC 2506  & 9.22 $\pm \ 0.175$ & 0.056 & -0.20 & 0.084 $\pm \ 0.006$ & 0.149 $\pm \ 0.009$ & 33 & 0.263 \\
    NGC 2420  & 9.24 $\pm \ 0.175$ & 0.013 & -0.38 & 0.079 $\pm \ 0.006$ & 0.14 $\pm \ 0.009$ & 7 & 0.210\\
    Ruprecht 68  & 9.26 $\pm \ 0.15$ & 0.446 & 0.13 & 0.151 $\pm \ 0.022$ & 0.240 $\pm \ 0.030$ & 9 & 0.221\\
    NGC 2627  & 9.27 $\pm \ 0.15$ & 0.139 & -0.12 & 0.066 $\pm \ 0.011 $ & 0.117 $\pm \ 0.015$ & 10 & 0.260\\
    Melotte 66 & 9.63 $\pm \ 0.15$ & 0.167 & -0.33 & 0.123 $\pm \ 0.007 $ & 0.217 $\pm \ 0.011$ & 20 & 0.262\\
    NGC 2682  & 9.63 $\pm \ 0.15$ & 0.067 & -0.102 & 0.103 $\pm \ 0.006$ & 0.170 $\pm 0.009$ & 7 & 0.622\\
    NGC 6791  & 9.80 $\pm \ 0.15$ & 0.157 & 0.32 & 0.161 $\pm \ 0.006 $ & 0.183	$\pm \ 0.009$ & 36 & 0.204\\ 
\hline
\end{tabular}
\end{table*}

\begin{table*}
\caption{SMC fields in circular annuli centered around clusters, where the cluster names are given in column 1. The age (log years) from fits to PARSEC \citep{2012MNRAS.427..127B} isochrones are applied to all fields, while the reddenings $E(B-V)$ are applied separately to each. The radius of each cluster (degrees) from \citet{2010A&A...517A..50G} is listed, and the aperture for field stars is an annulus from 1.5 times this radius to 5.0 times this radius. The number of red clump stars and the $d_{BP-RP}$ color difference, transformed from $d_{B-I}$, are listed. A metallicity of [Fe/H]=$-0.6$ \citep{2004AJ....127.1531H,2007A&A...472..101I} is assumed.}

\label{tab:SMC BP-RP}
\centering
\begin{tabular}{lcccccccc}
\hline
Name & Age  & E($B-V$) & [Fe/H] & $d_{BP-RP}$ & $d_{B-R}$ & N(RC) & Cluster radius \\
& (log years) & (mag) & (dex) & (mag) & (mag) & & (deg) & \\
\hline
    SMC 26 &   & 0.03 &  & 0.120 $\pm \ 0.007$ & 0.212 $\pm \ 0.010$ & 211 & 0.025 \\
    SMC 225 &  & 0.05 &  & 0.125 $\pm \ 0.010$ & 0.222 $\pm \ 0.016$ & 29 & 0.004 \\
    SMC 245 &  & 0.03 &  & 0.091 $\pm \ 0.008$ & 0.161 $\pm \ 0.013$ & 109 & 0.006 \\
    SMC 368 &  & 0.02 &  & 0.110 $\pm \ 0.009$ & 0.195 $\pm \ 0.014$ & 65 & 0.007 \\
    SMC 571 &  & 0.08 &  & 0.109 $\pm \ 0.007$ & 0.192 $\pm \ 0.010$ & 344 & 0.011 \\
    \hline
    Average & 8.85 $\pm \ 0.07$ & & -0.6 & 0.111 $\pm \ 0.004$ & 0.196 $\pm \ 0.007$ & \\
\hline
\end{tabular}
\end{table*}

\begin{figure}
	\includegraphics[width=\columnwidth]{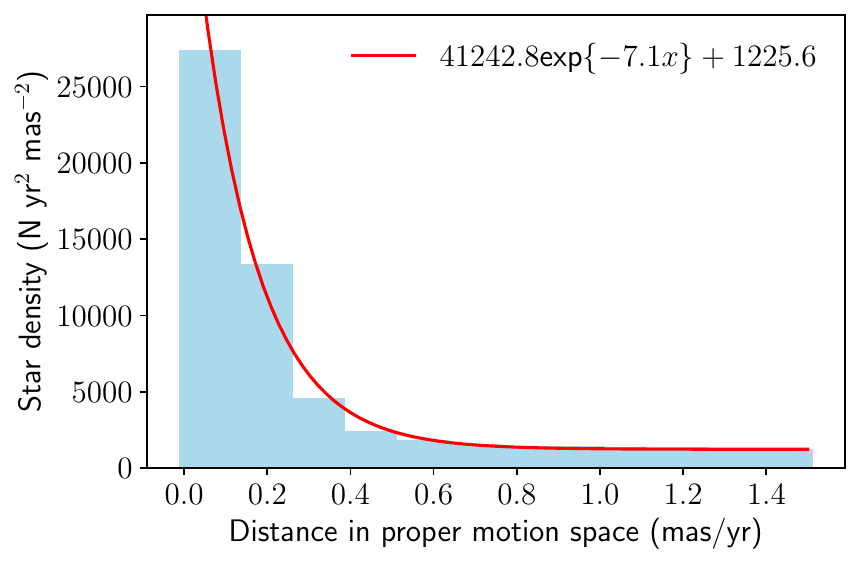}
    \caption{Stellar density in annuli proceeding from the center of the proper motion location of NGC~7789 amongst the field stars. An exponential model is shown (red). Field star density is taken as the constant-density asymptote.}
    \label{fig:pm_dist}
\end{figure}

\begin{figure}
	\includegraphics[width=\columnwidth]{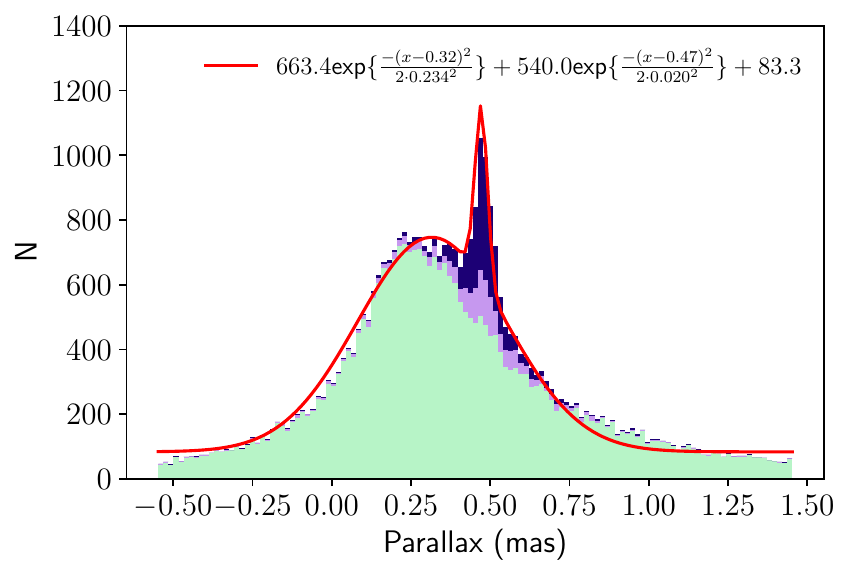}
    \caption{Distribution of stars in parallax in the NGC~7789 field. Stars nearest the proper motion locus (navy) and additional stars in a wider annulus (purple) form an obvious overdensity amongst the field stars (green). A two-Gaussian model is shown (red), with the broader component representing field stars.}
    \label{fig:plx_dist}
\end{figure}

The fit for proper motion assumed an exponential density profile for the cluster on top of a constant field density (Figure \ref{fig:pm_dist}), the only difference being that the ``distance'' is the $\Delta\mu$ from the cluster locus. The parallax distribution was fit with two Gaussians, a broad one for the field and a narrow one for the cluster (Figure \ref{fig:plx_dist}). Cone search opening angles (column 7 in Table \ref{tab:cluster BP-RP new}) were revisited if density profiles did not clearly reach an asymptote.

\begin{figure}
	\includegraphics[width=\columnwidth]{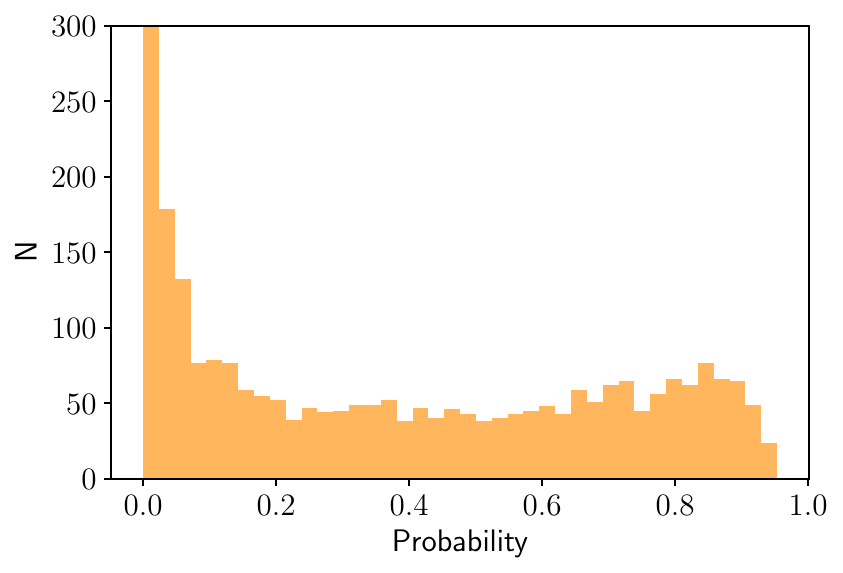}
    \caption{The resultant probability distribution after separate probabilities are combined for NGC~7789. The first bin contains about 20000 nonmember stars.}
    \label{fig:prob_dist}
\end{figure}

Given three separate probability estimates, the combined probability from separate $P_{(\alpha, \delta)}$, $P_\mu$, and $P_\pi$ estimates is computed as in \citet{1998A&AS..133..387B}, updated to include increased dimensionality as in \citet{2022MNRAS.511.4702G}. The resultant probability distribution for NGC 7789 is shown in Figure \ref{fig:prob_dist}.

% 2.2 - Empirical calculation
\section{Results} \label{section_results}

\begin{figure}
    \includegraphics[width=\columnwidth]{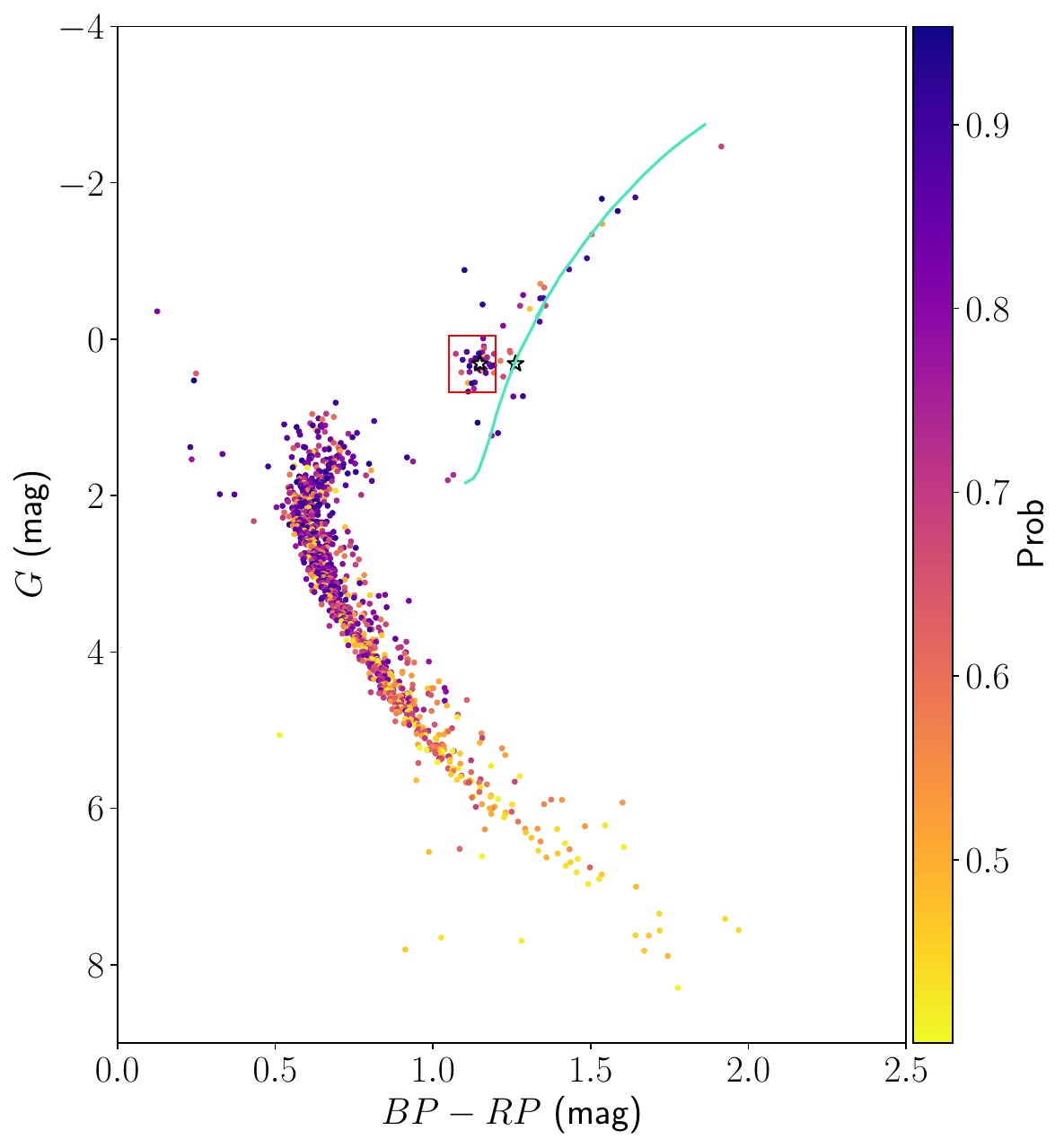}
    \caption{Color-absolute magnitude diagrams in \textit{Gaia} passbands for probable members ($P > 0.7$) of open cluster NGC~7789 using the distance and reddening from Table \ref{tab:cluster BP-RP new}. Symbols are color mapped by membership probability as indicated by the color bar. A rectangle (red) indicating the estimated boundary of the red clump, along with the fitted RGB segment of a PARSEC isochrone at [Fe/H] = 0.00 (turquoise), used to determine the median locations of the red clump and RGB are shown. The red clump and RGB loci are plotted as white stars with black outlines.}
    \label{fig:cmd}
\end{figure}

\begin{figure}
    \centering
        \includegraphics[width=0.95\columnwidth]{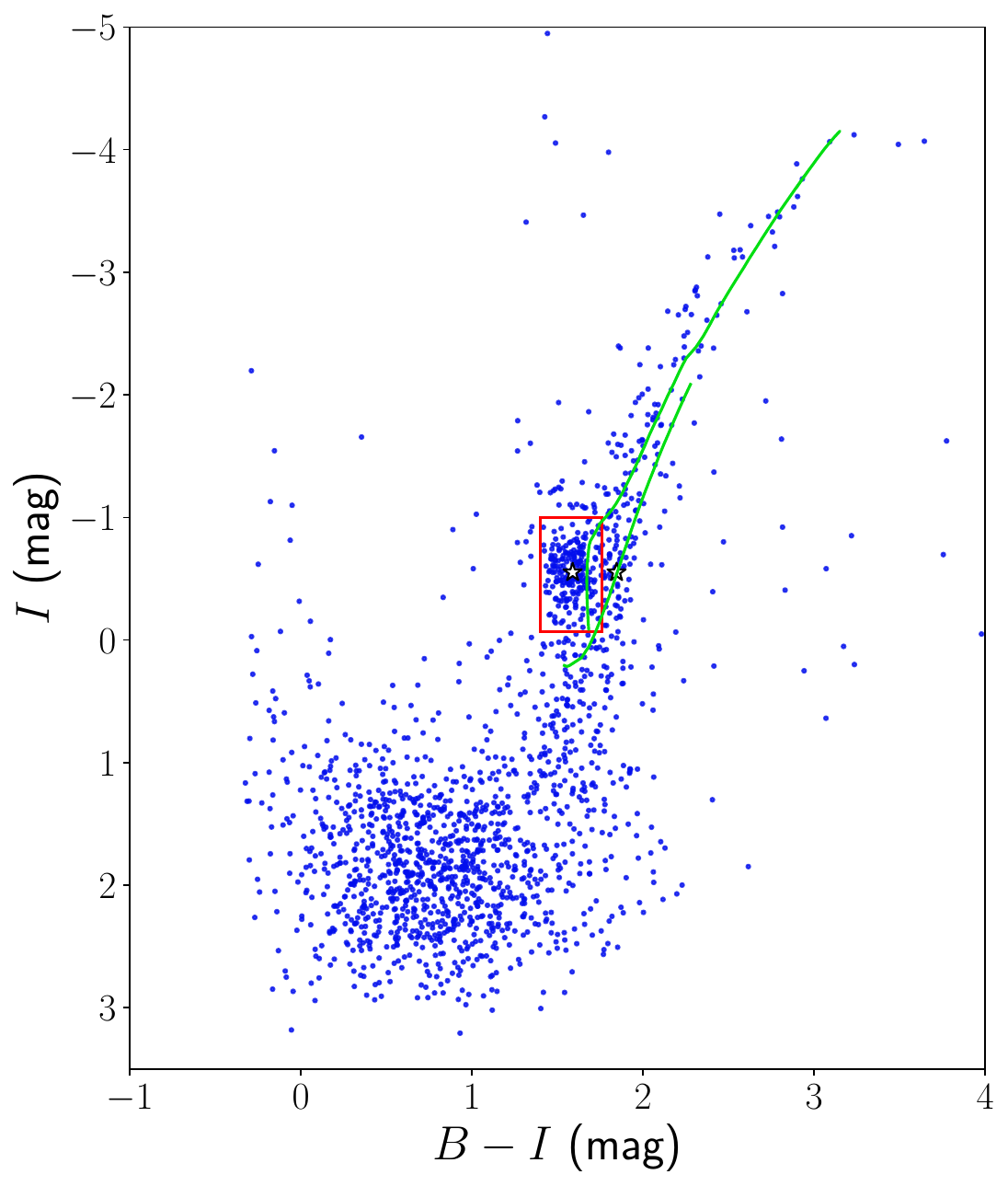}
    \caption{Color-absolute magnitude diagram of the field surrounding open cluster SMC 26, excluding the cluster itself. A rectangle (red) indicating the estimated boundary of the red clump, along with the fitted RGB segment a PARSEC isochrone at [M/H] $=-0.5$ (green), used to determine the median locations of the red clump and RGB, are shown. The red clump and RGB loci are plotted as white stars with black outlines.}
    \label{fig:smc_cmd}
\end{figure}

An example of a cleaned CMD appears in Figure \ref{fig:cmd} and the remainder of cluster CMDs appear in Appendix \ref{appendix1}. From the CMDs we drew boxes by hand and computed the central red clump color-magnitude location using the Tukey biweight statistic \citep{1990AJ....100...32B}, which is similar to the median, in both $BP-RP$ and $G$. The error in central location is estimated as $\sigma = \sqrt{v/N}$, where $v$ is the biweight midvariance and $N$ is the number of red clump stars used for the calculation. RGB colors at the $G$ magnitude of the red clump are found by overlaying the RGB segment of a PARSEC \citep{2012MNRAS.427..127B} isochrone on the data, using synthetic photometry from the Worthey models to generate the $BP-RP$ color. The synthetic RGB is then shifted left and right until fifty percent of stars were bluer and fifty percent of stars were redder. Once placed, the $BP-RP$ magnitude at the $G$ magnitude of the red clump is read from the shifted synthetic RGB. The clump stars outnumbered the RGB stars, so we expect counting statistics on the RGB to be the largest error source. We estimated RGB placement errors by shifting the theoretical RGB $\pm$ one star from the median position, and taking the average of the absolute value of those two shifts. For small $N$, our biggest concern, this is a good error estimate assuming the nearest neighbor distribution function from a Poisson point process. For large $N$, the method converges faster than $N^{-1/2}$ and become inappropriate, but there is also a theoretical error floor from differences in RGB tilts between different isochrone sets. We compared isochrones of 1 Gyr age and solar composition from \citet{1994ApJS...95..107W,2004ApJ...612..168P,2018ApJ...856..125H,1994ApJS...94...63B,2008A&A...482..883M,2012MNRAS.427..127B} and \citet{2007ApJ...666..403D} and performed statistics on the RGB slopes to arrive at rough error floors of 0.004 mag for $d(BP-RP)$ and 0.007 mag for $d(B-R)$. These errors were added in quadrature with the nearest neighbor errors for the final value tabulated in Tables \ref{tab:cluster BP-RP new} and \ref{tab:SMC BP-RP}.

Unsolved astrophysical effects and the placement of the initial red clump selection box may slightly skew the Tukey biweight for finding the central location. An example of an astrophysical effect occurs for clusters where the initial mass of a star at the RGB tip is $\sim 2$ M$_\odot$. NGC 7789 and NGC 752 (which is not in our sample) \citep{2000A&A...354..892G} have a clump with a few extra stars that are both lower luminosity and slightly blue. Significant spread in mass loss on the RGB combined with the fact that this mass might straddle the mass at which helium burning begins in a degenerate versus a nondegenerate core might explain the effect. Many of our clusters have turnoff masses near 2 M$_\odot$, and some (NGC 1245, NGC 2477) have ``tall'' red clumps not predicted by theory that could potentially be explained by this effect. On the other hand, theory is currently unable to account for fundamental red clump properties \citep{2000ASPC..211..169D,2019ApJ...879...81A}. The mass loss history, be it pre-flash or flash-caused, is frequently invoked as a possible explanation. A separate example at older ages but high metallicity is NGC 6791, some of whose RGB stars may skip helium burning altogether \citep{2005ApJ...635..522H}.

A ``tall'' clump also generates a new error source, namely uncertainty in the luminosity of clump. Transferred to the RGB, the luminosity uncertainty translates to a color uncertainty, albeit a mild one due to the steepness of the RGB. This error was included by computing a biweight location and scale in the $G$ dimension and scaling it appropriately via the $dG/d(BP-RP)$ RGB slope.

Additional youthful clusters in the SMC were investigated using the Magellanic Clouds Photometric Survey \citep{2002AJ....123..855Z} but separation of cluster and field proved ambiguous. However, some the fields surrounding many of the clusters appeared to be amenable to analysis, with a large fraction of the stars appearing to land at a single, common age. We assumed a metallicity of [Fe/H] $=-0.6$ \citep{2004AJ....127.1531H,2007A&A...472..101I} and fitted an age using PARSEC isochrones at an abundance value of [M/H] = $-0.5$ as summarized in Table \ref{tab:SMC BP-RP}. Figure \ref{fig:smc_cmd} illustrates CMDs for one of the five fields. The methods for calculating the median location of the red clump and the RGB color at the level of the red clump are the same as those described above for the Milky Way clusters. For the SMC CMDs, we estimated the error in the RGB color as $\delta = \sigma / \sqrt{N_{RGB}} = 0.2T_{RGB} / \sqrt{N_{RGB}}$, where $T_{RGB}$ is the thickness of the RGB at the level of the red clump and $N_{RGB}$ is the number of RGB stars used to compute the median location. We estimated $T_{RBG}$ as two times the color difference between the median and reddest RGB star. We assume that the spread in $BP-RP$ of the RGB follows a normal distribution, so 99.38\% of RGB stars at a given $G$ magnitude are within 2.5 standard deviations of the median location. Thus $2 \cdot 2.5 \sigma = T_{RGB}$.

For Milky Way clusters, we adopted ages and rough errors from \citet{2020A&A...640A...1C}, namely $\pm$ 0.175 log years for clusters younger than 9.25 log years and $\pm$ 0.15 log years for older clusters. To plot \citetalias{1991MNRAS.251..545H}'s data, we use ages and age error estimates from \citet{2013A&A...558A..53K} for globular clusters Pal 12 and 47 Tuc and \citet{2012ApJ...754..108B} for globular clusters Pal 4 and Eridanus. For SMC clusters, we use \citet{1998AJ....115.1934D} and \citet{1998AJ....116.2395M}. For LMC clusters, we use \citet{1995A&A...298..427B} for NGC 1978, \citet{2007A&A...462..139K} for NGC 2173, \citet{10.1093/mnras/stw1260} for Hodge 4 and ESO121-SC03, and \citet{1991AJ....101..515O} for LW 47. We transform the \citetalias{1991MNRAS.251..545H} errors in age from years to log years using $\delta(\log(t)) = \delta(t)/(t\ln 10)$.

We took metallicities from \citet{2012ApJ...754..108B} for Eridanus, \citet{2013A&A...558A..53K} for Pal 12 and 47 Tuc, \citet{1996AJ....112.1487H} for Pal 4, \citet{2007A&A...462..139K} for NGC 2173, and \citet{10.1093/mnras/stw1260} for Hodge 4 and ESO121-SC03. For all other \citetalias{1991MNRAS.251..545H} clusters, we used the metallicity listed therein.

We used the algebraic formulae in \citet{2021A&A...649A...3R} to transform from the \textit{Gaia} photometric system to Johnson-Cousins $B-R$. Color excess from \cite{2013A&A...558A..53K} was converted from E(B-V) to E($BP-RP$) using \citet{2018MNRAS.479L.102C}. 

\begin{figure*}
    \centering
    \includegraphics[width=\textwidth]{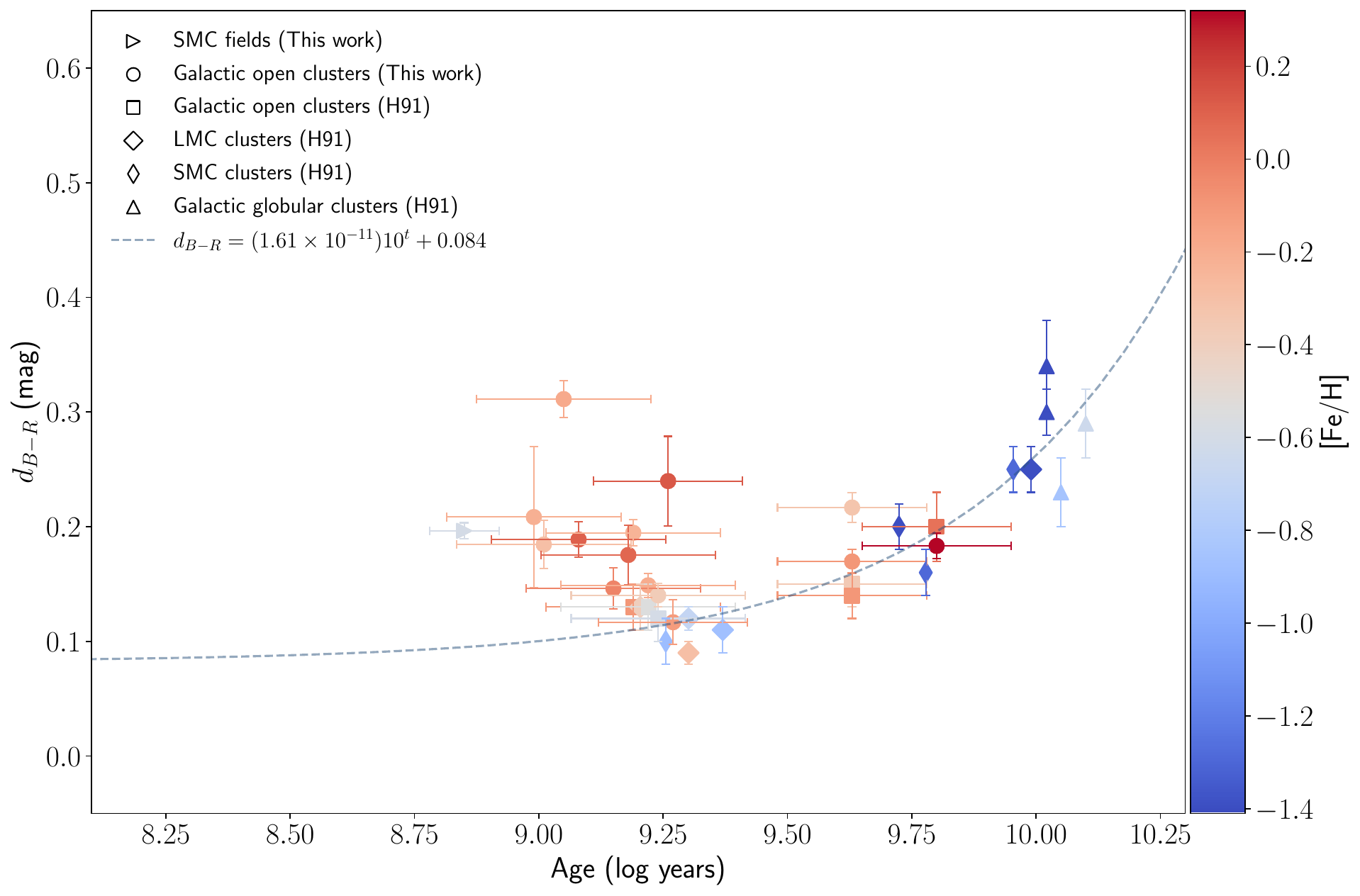}
    
    \caption{Color delta versus log age for galactic open clusters (circles) and average of SMC fields (right-facing triangles) from this work, along with galactic open clusters (squares), galactic globular clusters (pointing-up triangles), SMC clusters (thin diamonds), and LMC clusters (wide diamonds) from \citetalias{1991MNRAS.251..545H} color mapped by metallicity as indicated by the color bar. The \citetalias{1991MNRAS.251..545H} linear fit (dashed line) is also shown.}
    
    \label{fig:age_calibration}
\end{figure*}

Results are summarized in Table \ref{tab:cluster BP-RP new} and shown in Figure \ref{fig:age_calibration} with the \citetalias{1991MNRAS.251..545H} fit, transformed to logarithmic age, overlaid.

\section{Discussion and Conclusion} \label{section_discussion}

From Figure \ref{fig:age_calibration} we find that the \citetalias{1991MNRAS.251..545H} line extrapolates poorly at the young end. At log age = 9.25, or age = 1.8 Gyr, the two samples overlap in age, but every cluster in the new sample (this work) shows a wider $d_{B-R}$ color separation. The magnitude of the effect exceeds any possible random or systematic uncertainties. The rise toward younger ages between log age $\sim$9.25 and $\sim$9.0 means that going from $d_{B-R}$ to age is double valued: One would need sufficient main sequence photometry to determine if the cluster was older or younger than $\sim$2 billion years before using either the old age \citetalias{1991MNRAS.251..545H} line or a young age line with a slope of opposite sign.

Even restricted to the new measurements, and further restricted to log age $< 9.5$, the scatter appears to be astrophysical. A Kolmogorov-Smirnov test on our $d_{BP-RP}$ values for clusters in our sample younger than 9.5 log years returned $p \approx 0.002$, indicating that the scatter is not drawn from a normal distribution. The cause of the scatter cannot be age, or we would see a slope in Figure \ref{fig:age_calibration}. We sample metallicity span less well, but within that caveat, no metallicity dependence is evident. If neither age nor metallicity is the cause of the modulation in $d_{B-R}$, we are driven to consider more subtle causes such as variation in helium abundance, variation in the [$\alpha$/Fe] ratio, or variation in binary separation  distributions. The investigation into the cause of the scatter must await future work.

\begin{figure*}
    \centering
    \includegraphics[width=\textwidth]{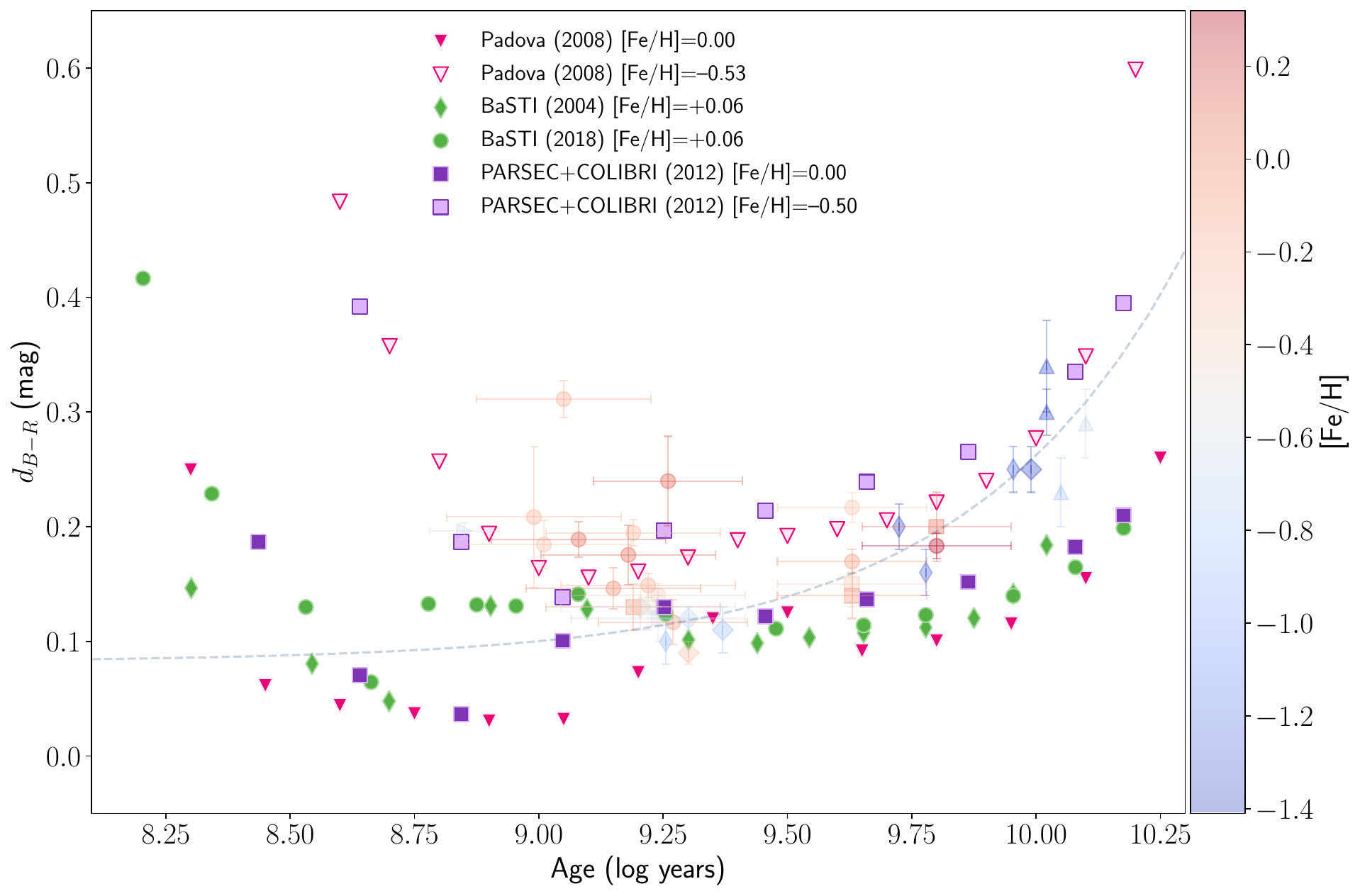}
    \caption{Color delta versus log age. Symbols are the same as Figure \ref{fig:age_calibration} with the addition of predictions from various isochrone sets: We show \citep{2008A&A...482..883M} at two abundance values (pink triangles, unfilled for the metal poor sequence), BaSTI 5.0.1 \citep{2004ApJ...612..168P} (green diamonds), BaSTI \citep{2018ApJ...856..125H} (green circles), and PARSEC+COLIBRI \citep{2012MNRAS.427..127B, 2020MNRAS.498.3283P} at two abundance values (purple squares, unfilled for the metal poor sequence). }
    \label{fig:theory_age_calibration}
\end{figure*}

We examine some predictions from stellar evolution theory in Figure \ref{fig:theory_age_calibration}. All isochrones were translated from $T_{\rm eff}$ and log~$L$ to $B-R$ using the empirical transformations of \citet{2011ApJS..193....1W}. We show BaSTI 5.0.1 \citep{2004ApJ...612..168P}, BaSTI updated \citep{2018ApJ...856..125H}, Padova, retrieved 2011 \citep{2008A&A...482..883M}, at two abundance values, and PARSEC+COLIBRI \citep{2012MNRAS.427..127B, 2020MNRAS.498.3283P} with Riemers \citep{1975MSRSL...8..369R} $\eta = 0.2$, retrieved 2023, also at two abundance values. Within the isochrone files, red clumps were located by treating the isochrone as a stellar population, predicting relative numbers of stars, and choosing the dominant sampling point to correspond to the red clump location. On the positive side, all evolutionary sets are within $\sim$0.1 mag (about 150 K) of the observed color delta, and all sets show a growth in $d_{B-R}$ for ancient stellar populations that qualitatively mimics the observations.

If precision better than 150 K is desired, theory does not match observation. In particular, reading from Figure \ref{fig:theory_age_calibration}, theory predicts a metallicity dependence of roughly $\frac{d(B-R)}{d{\rm [Fe/H]}} \geq -0.2$ mag per decade in heavy element abundance. A metallicity dependence this large is strictly ruled out by the data, which includes clusters from the large and small Magellanic Clouds. For example, super-metal-rich cluster NGC~6791 lies right among SMC clusters. The presence of this systematic among current isochrone sets affects CMD decompositions \citep{2001ApJS..136...25H, 2002MNRAS.332...91D}. It also impacts integrated light models by introducing a fictitious color drift in the sense to make metal rich populations redder.

To estimate the size of the induced integrated light color change, we employ evolutionary population synthesis models \citep{2022MNRAS.511.3198W}. As of this writing, these models have eight options for stellar evolution, but the earliest one \citep{1994ApJS...95..107W} uses the \citetalias{1991MNRAS.251..545H} scheme to place the red clump. We added an extra line of code to add a temperature shift, and translated the models that span metallicity in Figure \ref{fig:theory_age_calibration} from $\Delta$ color to $\Delta$ temperature via \cite{2011ApJS..193....1W}. The results are summarized in Table \ref{tab:integratedlight}. Because the main sequence dominates in the blue, $U-V$ is affected very little, but there is about a $4\%$ effect in $V-K$. 

Modelers should be aware of this discrepancy between current stellar evolution theory and observation because it adds to a list of ill-modeled or un-modeled effects that are far larger than the RMS fit between galaxy spectra and modeled integrated light spectra. \cite{2014ApJ...780...33C}, for example, find fits as good as $0.2\%$ RMS, but their red clump temperatures do not adhere to \citetalias{1991MNRAS.251..545H} and so could be off by several percent. Blue stragglers are not included, which can lead to errors of over a magnitude in the terrestrial UV \citep{1999ApJ...524..824D}. Chemically peculiar stars, not included, alter the blue spectrum by $\sim 2\%$ in the blue \citep{2023MNRAS.518.4106W}. Binary evolution products, carbon stars, and metallicity-composite populations are likewise not modeled, and yet the optical spectrum matches to $0.2\%$. The reason the fit is excellent is a generalization the age-metallicity degeneracy \citep{1994ApJS...95..107W}, where instead of age or metallicity, one inserts the concept of stellar temperatures in general. While some astrophysical inferences, such as abundance ratios, are somewhat isolated from this degeneracy, others, such as age, metallicity, age spread, or metallicity spread are not.

\begin{table}
	\centering
	\caption{Integrated light systematic color error if theoretical $Z$ dependence is allowed to flourish.}
	\label{tab:integratedlight}
 \begin{tabular}{llll}
		\hline
  Quantity & log age   & log age  & log age  \\
           &  = 9  &  = 9.5 &  = 10 \\
		\hline
  $\Delta T_{\rm eff}$ & 172 K & 93 K & 212 K \\
  \hline
$U-V$, MP, H91 &  0.75 & 1.11 & 1.43 \\
$U-V$, MR, H91 &  1.04 & 1.58 & 1.99 \\
$U-V$, MR, $+\Delta$theory & 1.04 & 1.58 & 1.99 \\
Excess $U-V$ & 0.004 & 0.002 & 0.003 \\
	\hline
$V-K$, MP, H91 & 2.39 & 2.73 & 3.04 \\
$V-K$, MR, H91 & 2.70 & 3.33 & 3.81 \\
$V-K$, MR, $+\Delta$theory & 2.74 & 3.35 & 3.85 \\
Excess $V-K$ & 0.038 & 0.021 & 0.037 \\
  		\hline
	\end{tabular}
\end{table}

Future work on red clump systematics might include cross matching \textit{Gaia}'s star list with photometric catalogs with accuracies better than \textit{Gaia}'s $BP-RP$. The costs might include possible loss of stars from the sample and also possible exposure to systematic error caused by transforming from heterogeneous photometric systems to a common one. To target more clusters might also be contemplated, but the key is to cover parameter space, and the present combined data of \citetalias{1991MNRAS.251..545H} and ourselves covers well what the Milky Way provides.

\begin{figure}
    \centering
    \includegraphics[width=\columnwidth]{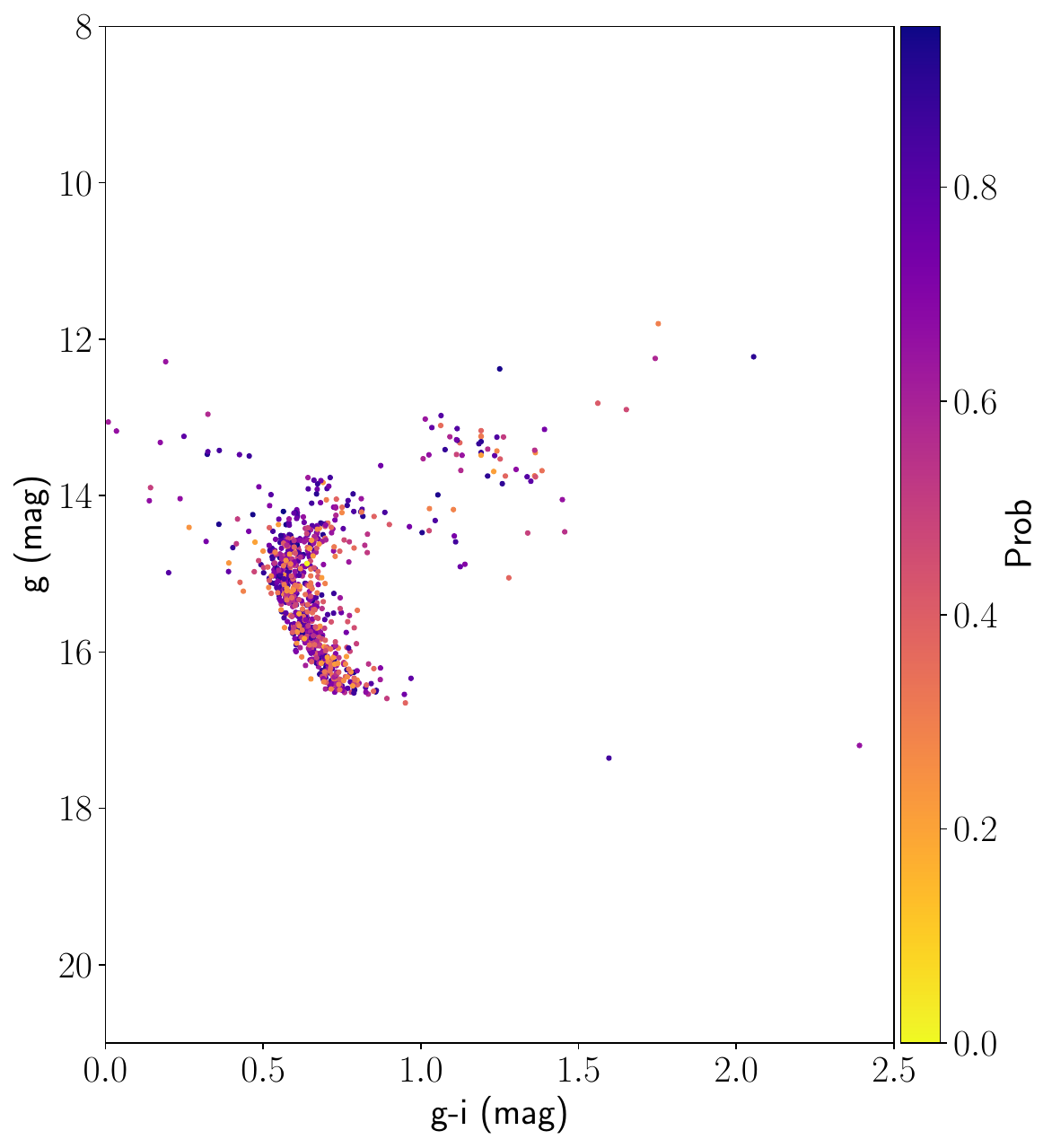}
    \caption{Color-apparent magnitude diagram of NGC 7789 from PanSTARRS $g$ and $i$ passbands. Points are colored by \textit{Gaia} membership probability as indicated by the color bar. Stars with apparent \textit{Gaia} $G$ magnitudes greater than 16 mag were not included in the cross-match so are not plotted.}
    \label{fig:panstarrs_cmd_ngc7789}
\end{figure}

We checked Milky Way clusters in this study that were also included in the PanSTARRS catalog to see if PanSTARRS photometry would clarify the CMDs and increase the accuracy of red clump or RGB color measurement. We cross matched \textit{Gaia} and PanSTARRS photometry for stars with \textit{Gaia}-estimated membership probability greater than 20\% and brighter than $G = 16$ mag. However, we found no improvement in red clump and RGB statistics. An example is shown in Figure \ref{fig:panstarrs_cmd_ngc7789}.

There are young-age limits for the $d_{B-R}$ method. For ages less than $\sim$150 Myr (log age 8.2), or, equivalently, MSTO masses greater than $\sim$5 M$_\odot$, stellar evolution through the RGB tip sparks no helium flash, and the helium-burning ``blue plumes'' no longer resemble red clumps. But there is a softer, practical limit that sets in around 600 Myr (log age 8.8) in which the RGB tip becomes fainter with youth, nearing the red clump luminosity. This decreases the number of first-ascent giants available for counting and increases uncertainty.

\section*{Acknowledgements}
        A.C. gratefully acknowledges the support of the Research Experiences for Undergraduates program, sponsored by the National Science Foundation Division of Physics Grant \#2050866. The WSU Department of Physics and Astronomy provided additional support. This work has made use of data from the European Space Agency (ESA) mission \textit{Gaia} (\url{https://www.cosmos.esa.int/gaia}), processed by the {\it Gaia} Data Processing and Analysis Consortium (DPAC, \url{https://www.cosmos.esa.int/web/gaia/dpac/consortium}). Funding for the DPAC has been provided by national institutions, in particular the institutions participating in the \textit{Gaia} Multilateral Agreement. This research also made use of the SIMBAD database, operated at CDS, Strasbourg, France.

\vspace{5mm}
%\facilities{HST(STIS), Swift(XRT and UVOT), AAVSO, CTIO:1.3m, CTIO:1.5m,CXO}

%% Similar to \facility{}, there is the optional \software command to allow 
%% authors a place to specify which programs were used during the creation of 
%% the manuscript. Authors should list each code and include either a
%% citation or url to the code inside ()s when available.

\software{{Astropy \citep{astropy:2013, astropy:2018, astropy:2022}, SciPy \citep{2020SciPy-NMeth}}}

%% Appendix material should be preceded with a single \appendix command.
%% There should be a \section command for each appendix. Mark appendix
%% subsections with the same markup you use in the main body of the paper.

%% Each Appendix (indicated with \section) will be lettered A, B, C, etc.
%% The equation counter will reset when it encounters the \appendix
%% command and will number appendix equations (A1), (A2), etc. The
%% Figure and Table counter will not reset.

\appendix

\section{Color Magnitude Diagrams \label{appendix1}}

\begin{figure}[hp]

    \gridline{ \fig{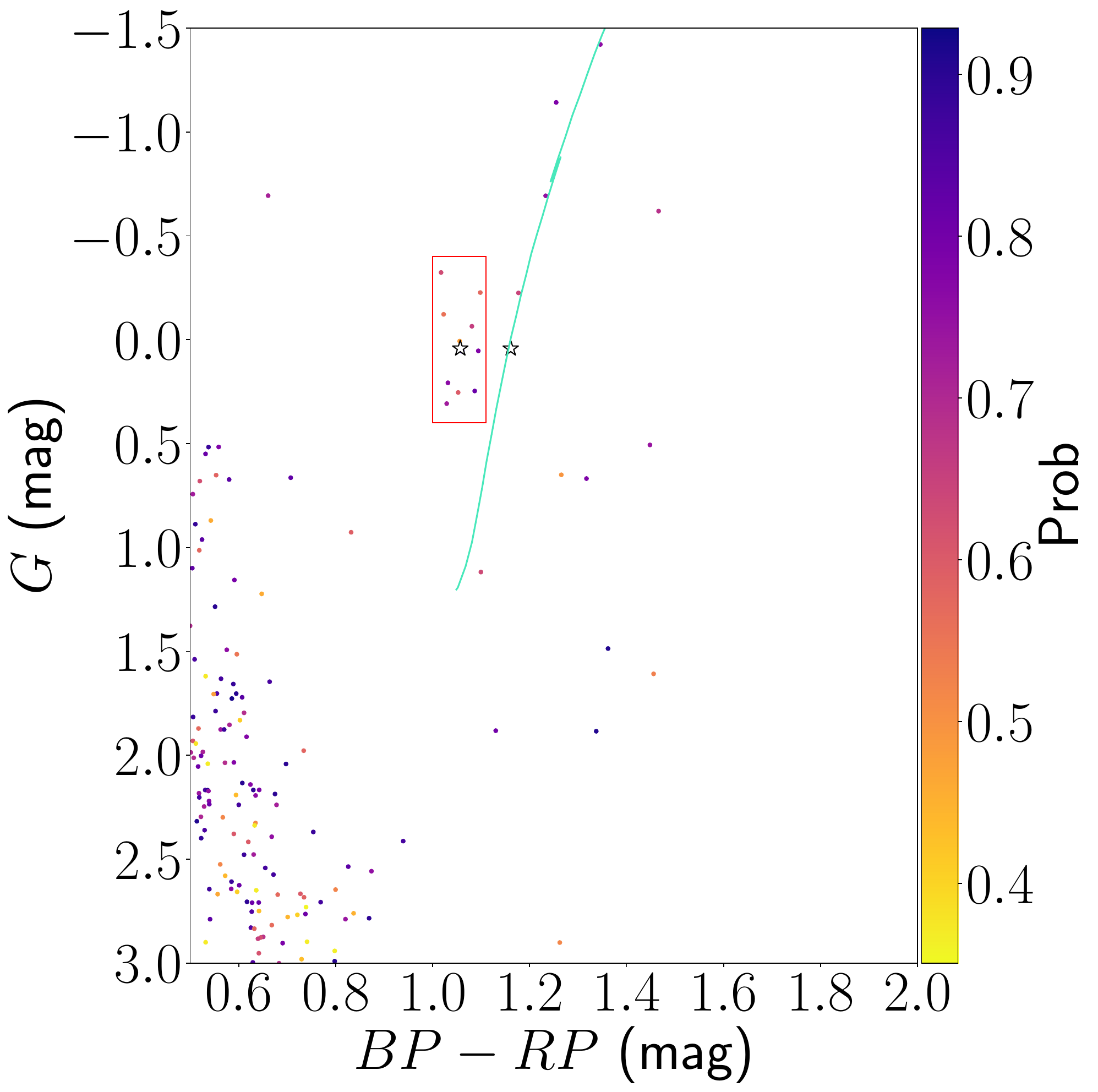}{0.3\textwidth}{(a) King~5}
      \fig{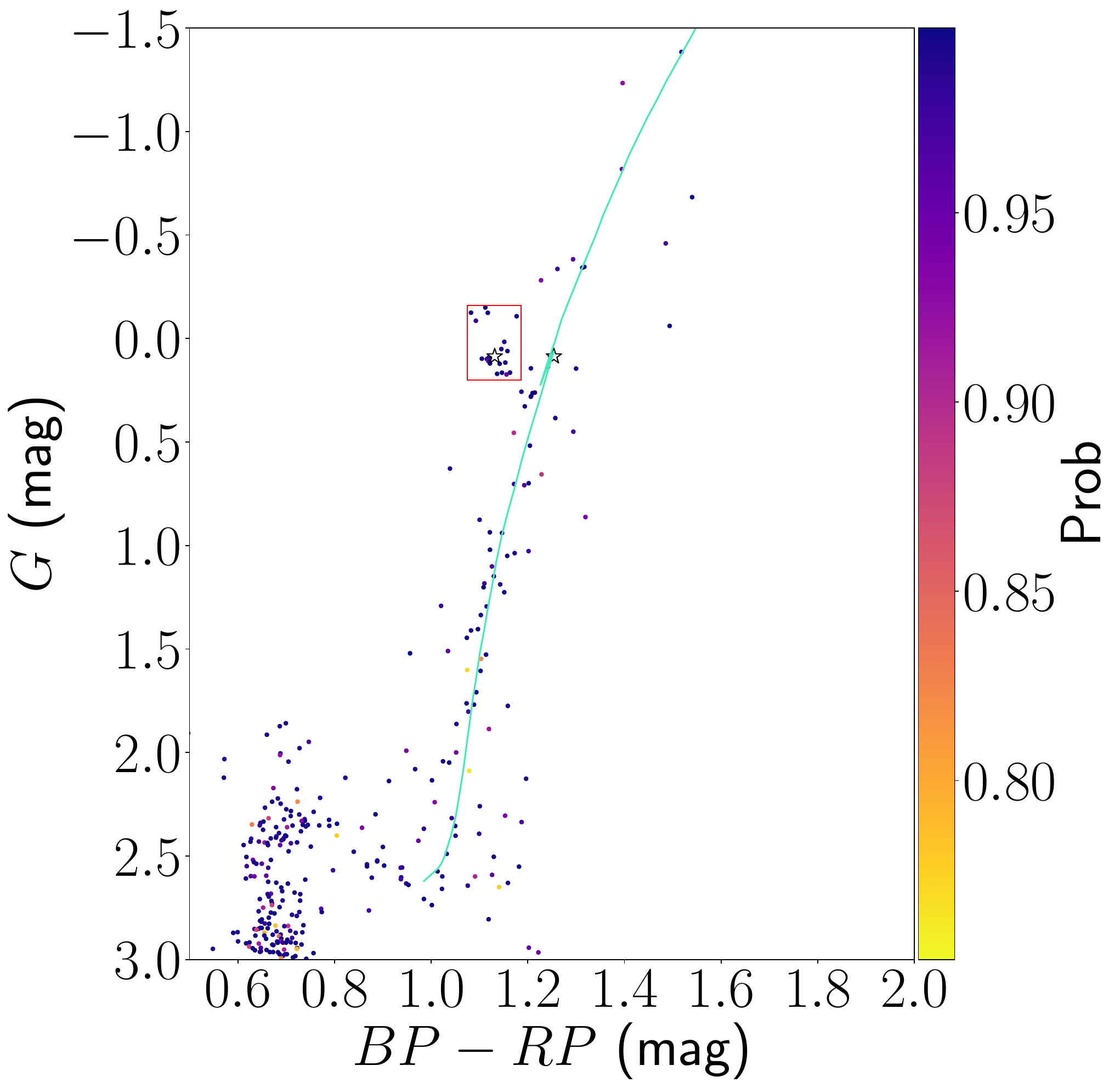}{0.306\textwidth}{(b) Melotte~66.}
      \fig{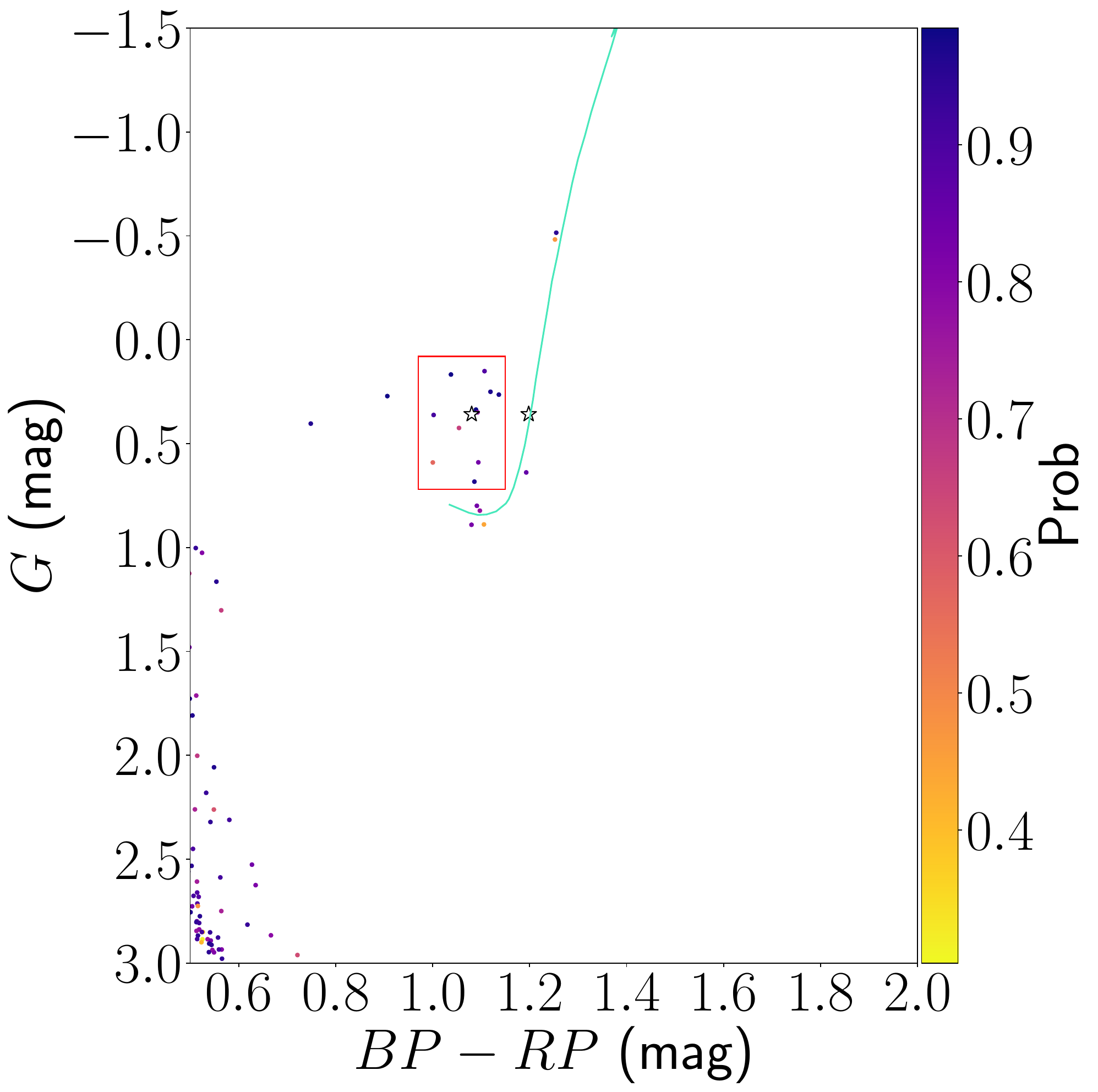}{0.3\textwidth}{(c) Melotte~71.}
    }
    \gridline{ \fig{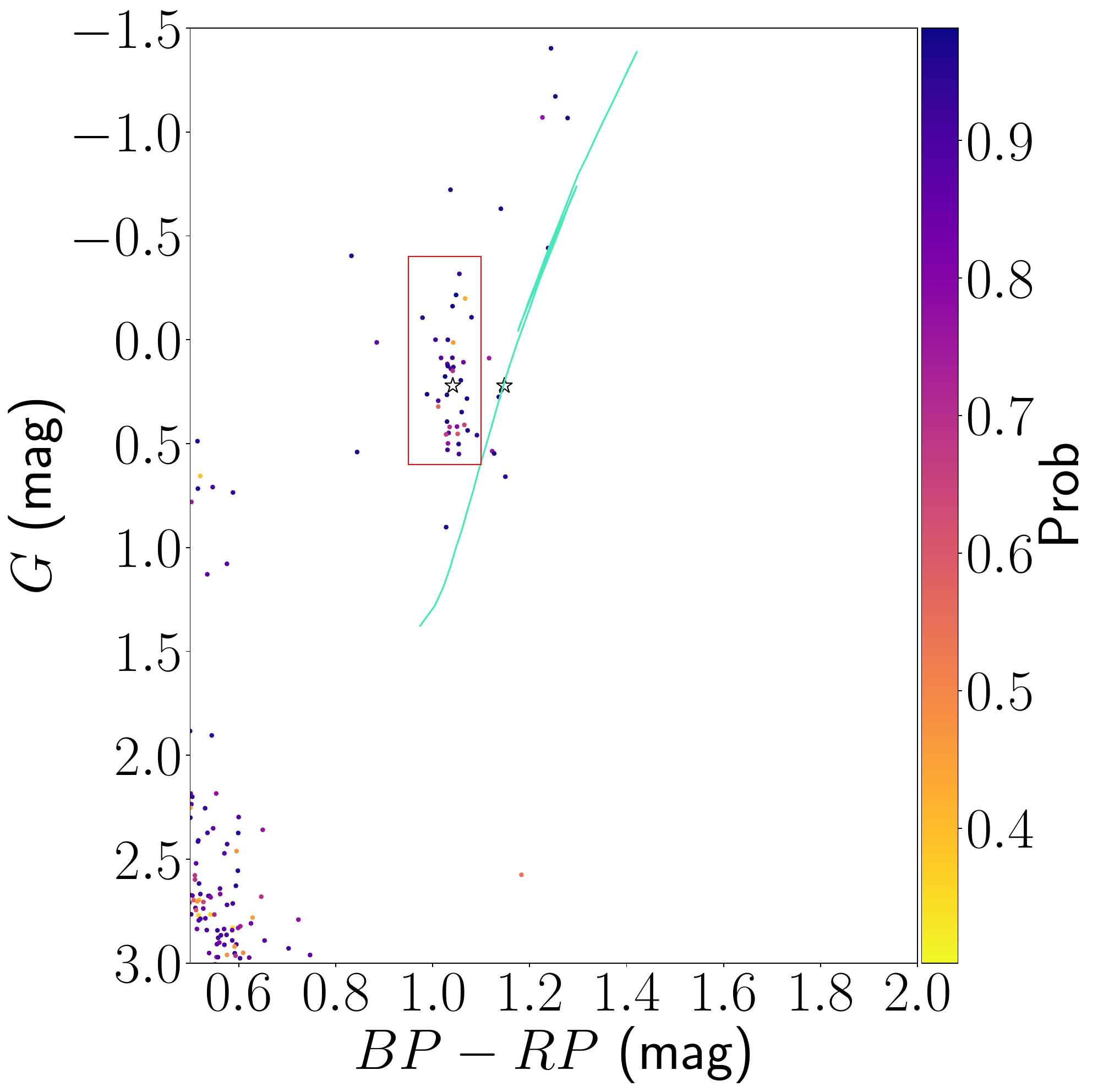}{0.3\textwidth}{(d) NGC~1245.}
      \fig{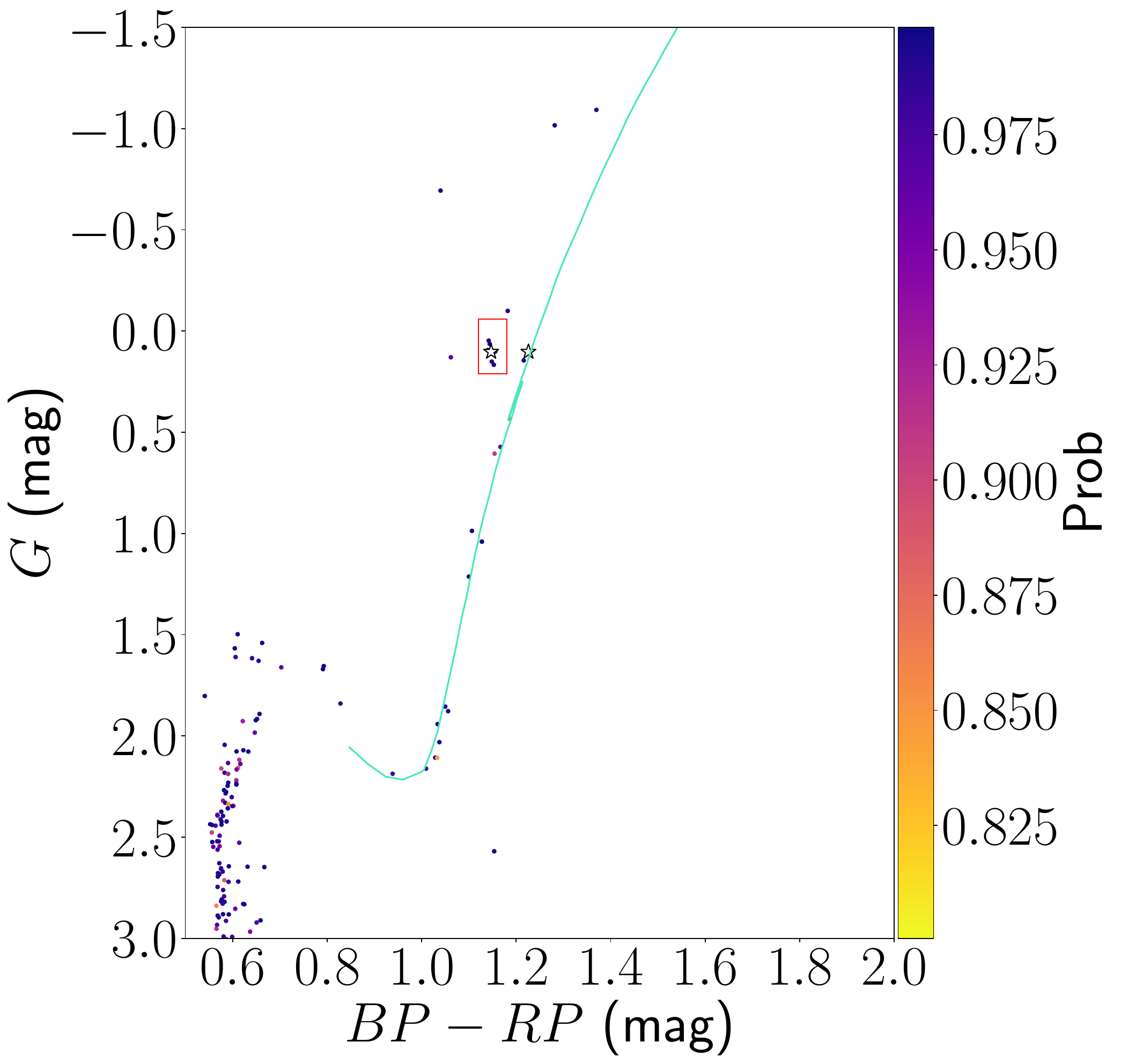}{0.315\textwidth}{(e) NGC~2420.}
      \fig{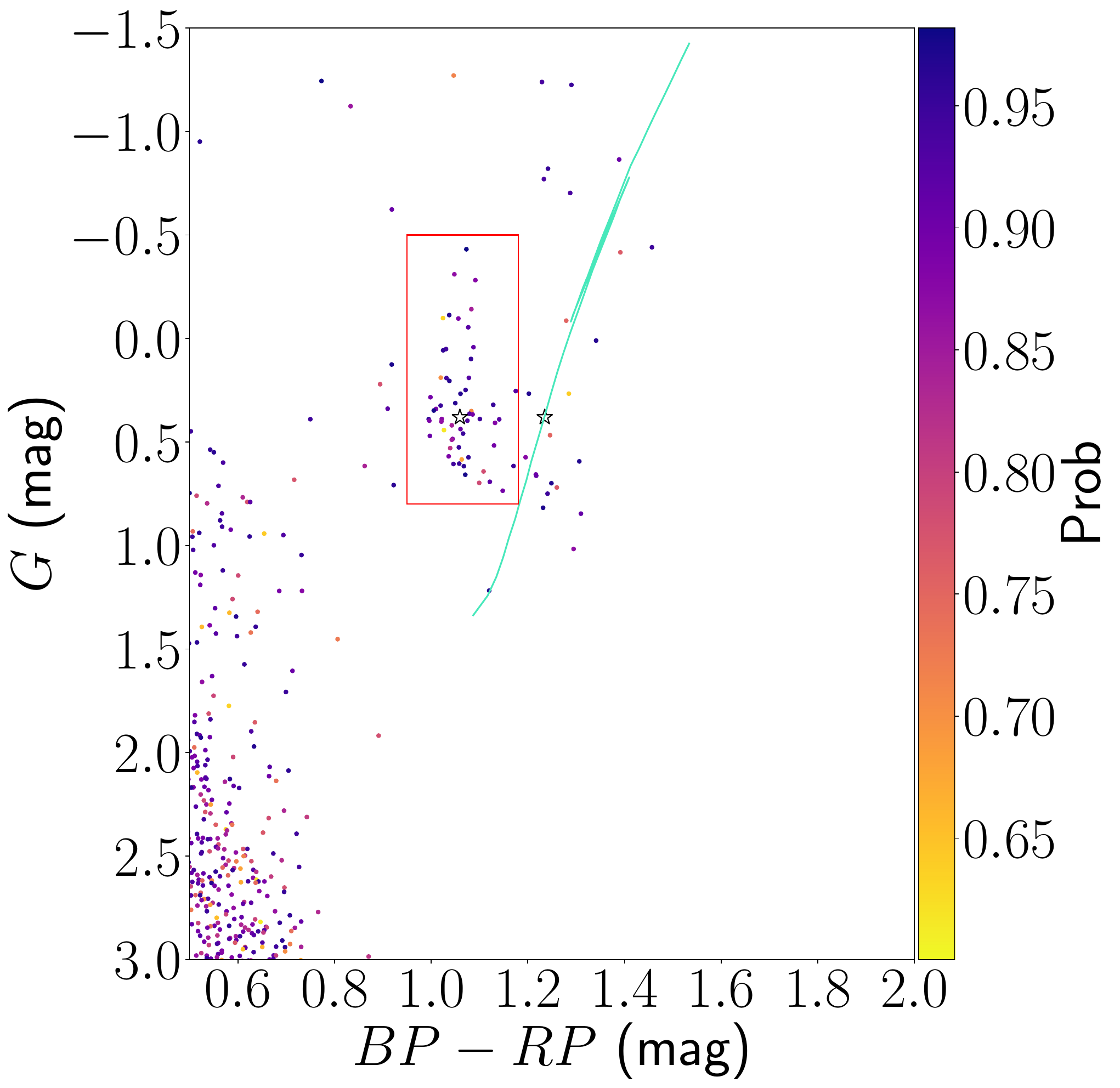}{0.31\textwidth}{(f) NGC~2477.}
    }
    \gridline{ \fig{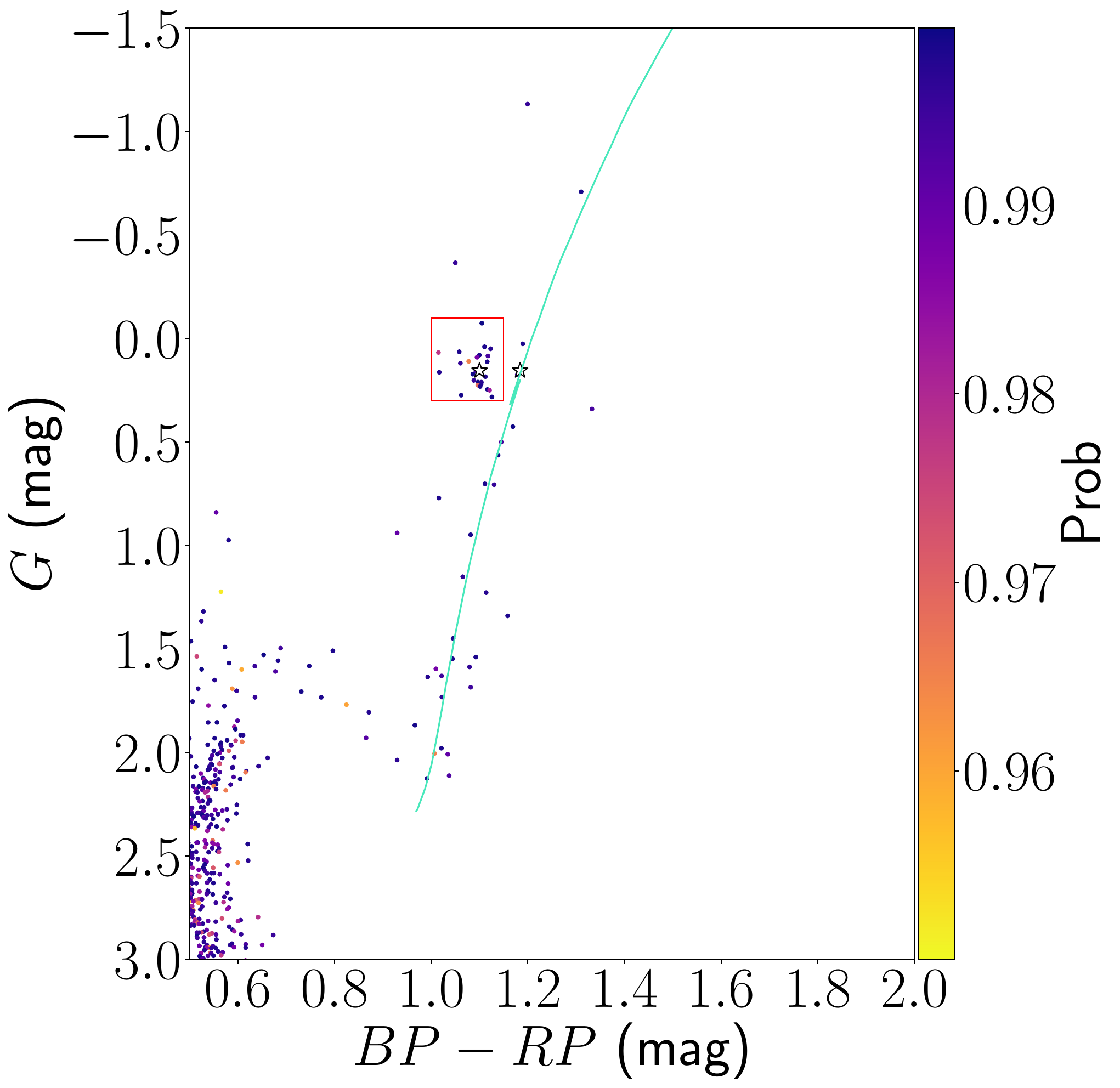}{0.307\textwidth}{(g) NGC~2506.}
      \fig{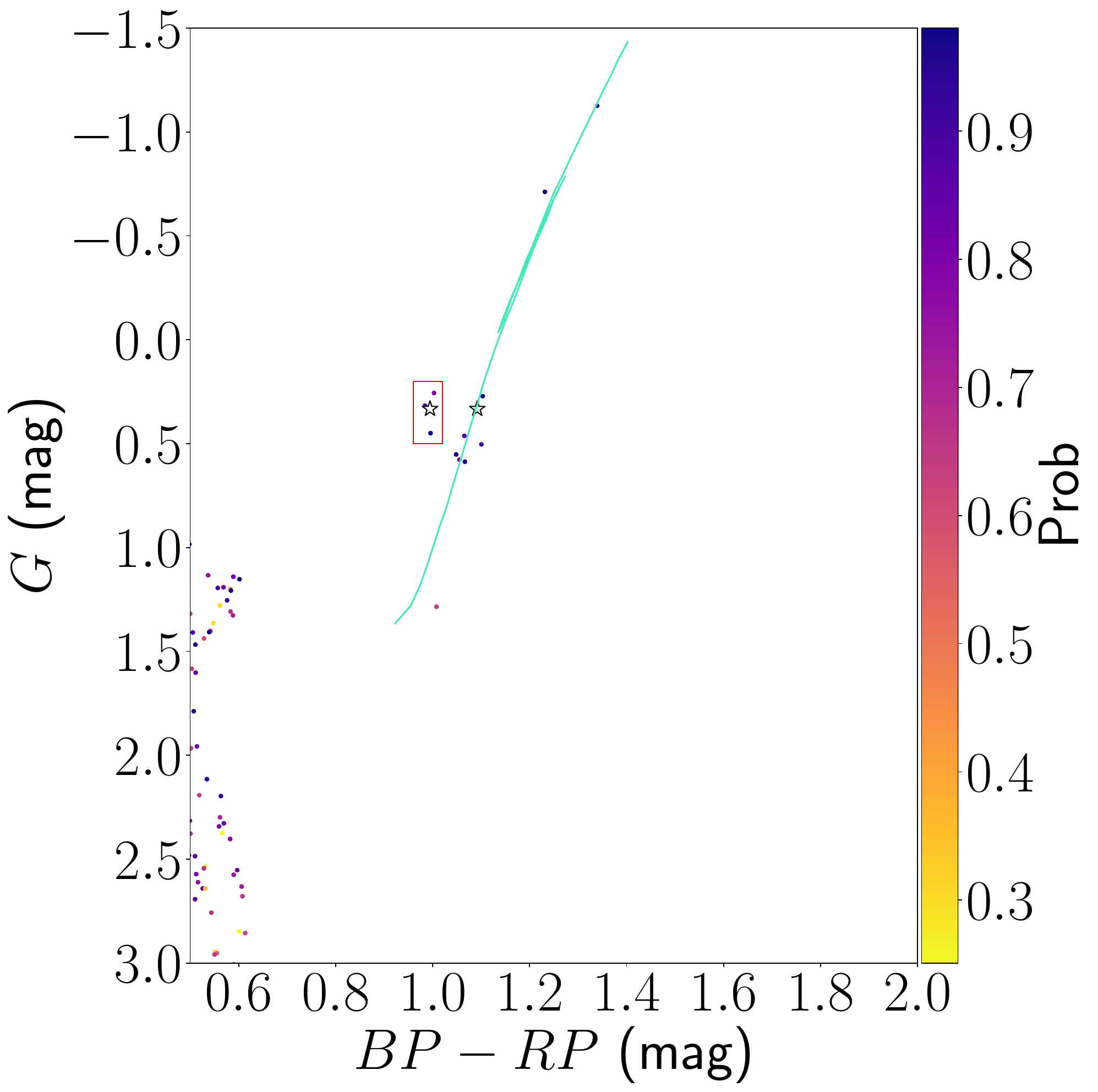}{0.3\textwidth}{(h) NGC~2509.}
      \fig{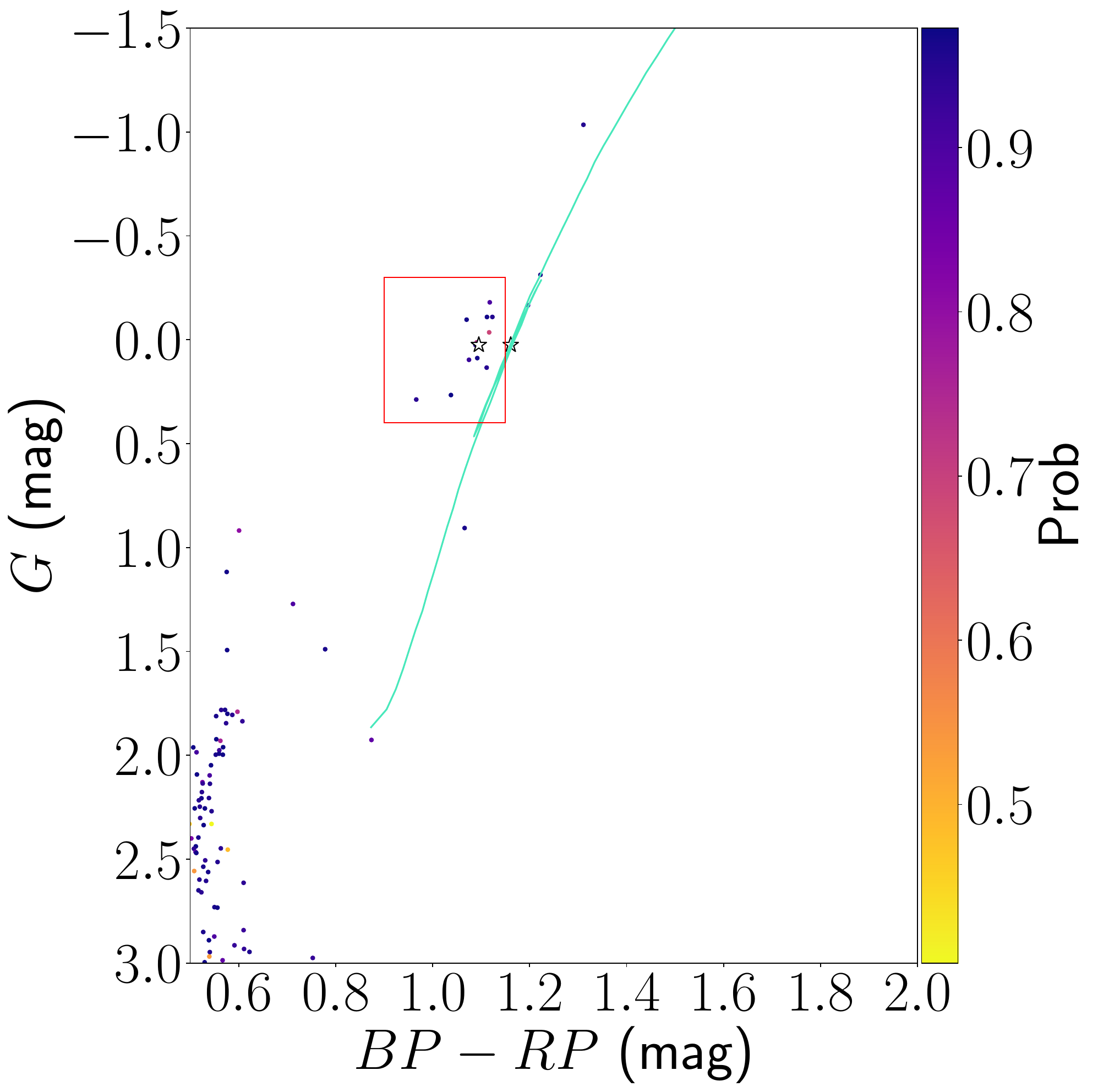}{0.3\textwidth}{(i) NGC~2627.}
    }

\caption{Color-absolute magnitude diagrams in \textit{Gaia} passbands for probable members of several Milky Way open clusters using the distance and reddening from Table \ref{tab:cluster BP-RP new}. Symbols and colors are the same as in Figure \ref{fig:cmd}.}
\label{fig:rc_rgb_cmd_full}
\end{figure}

\begin{figure}
    \gridline{ \fig{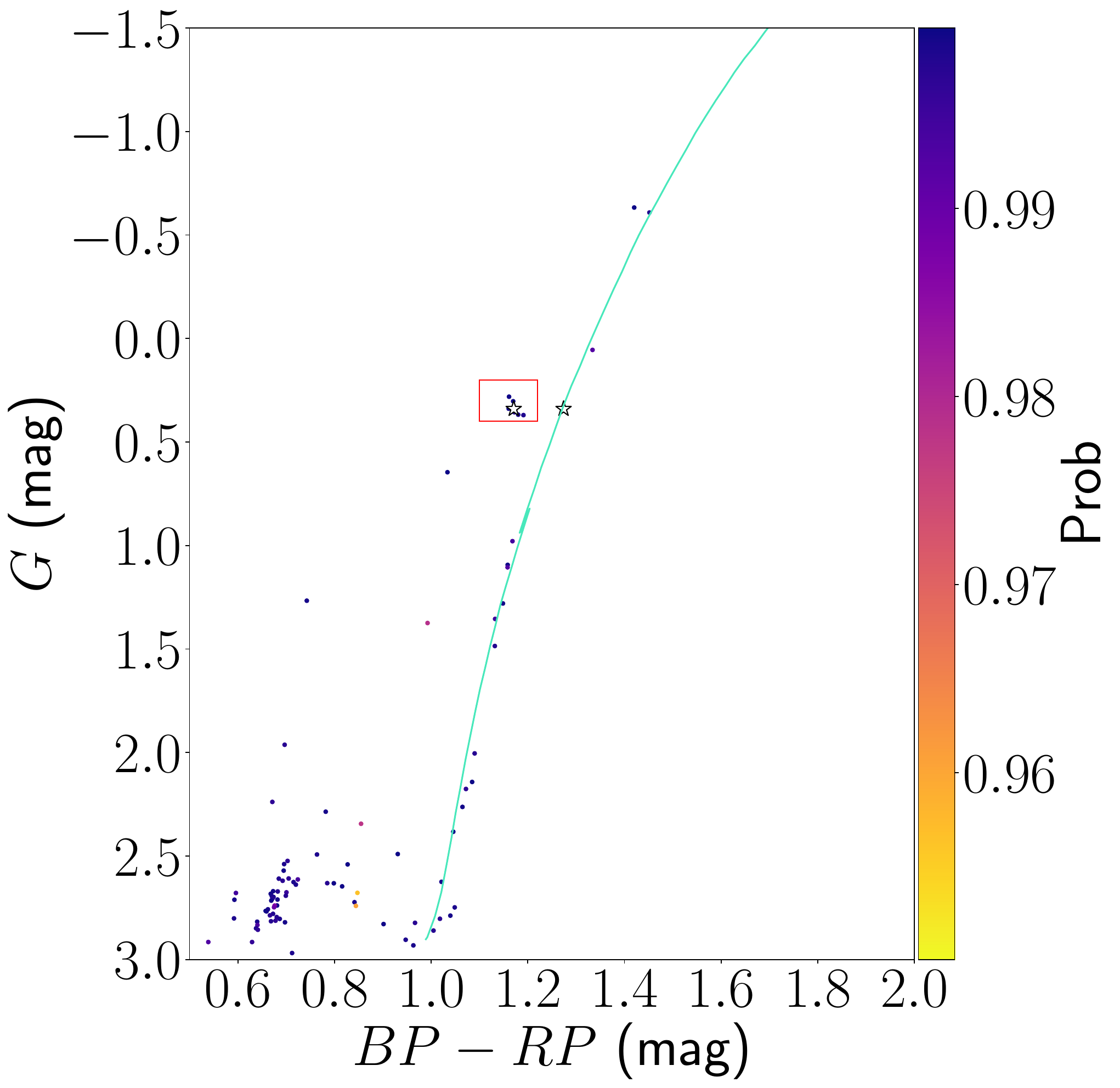}{0.31\textwidth}{(j) NGC~2682.}
      \fig{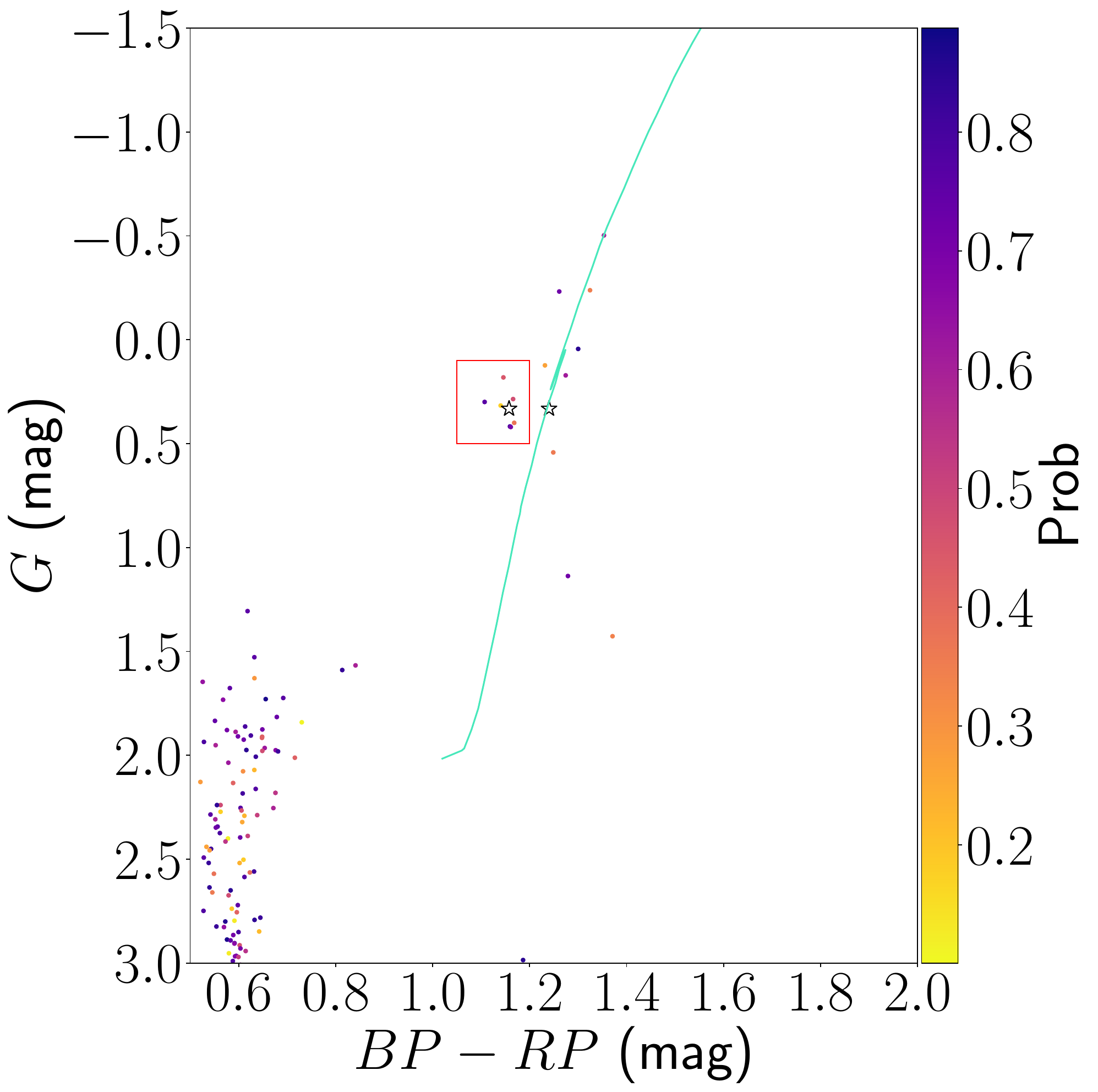}{0.302\textwidth}{(k) NGC~6208.}
      \fig{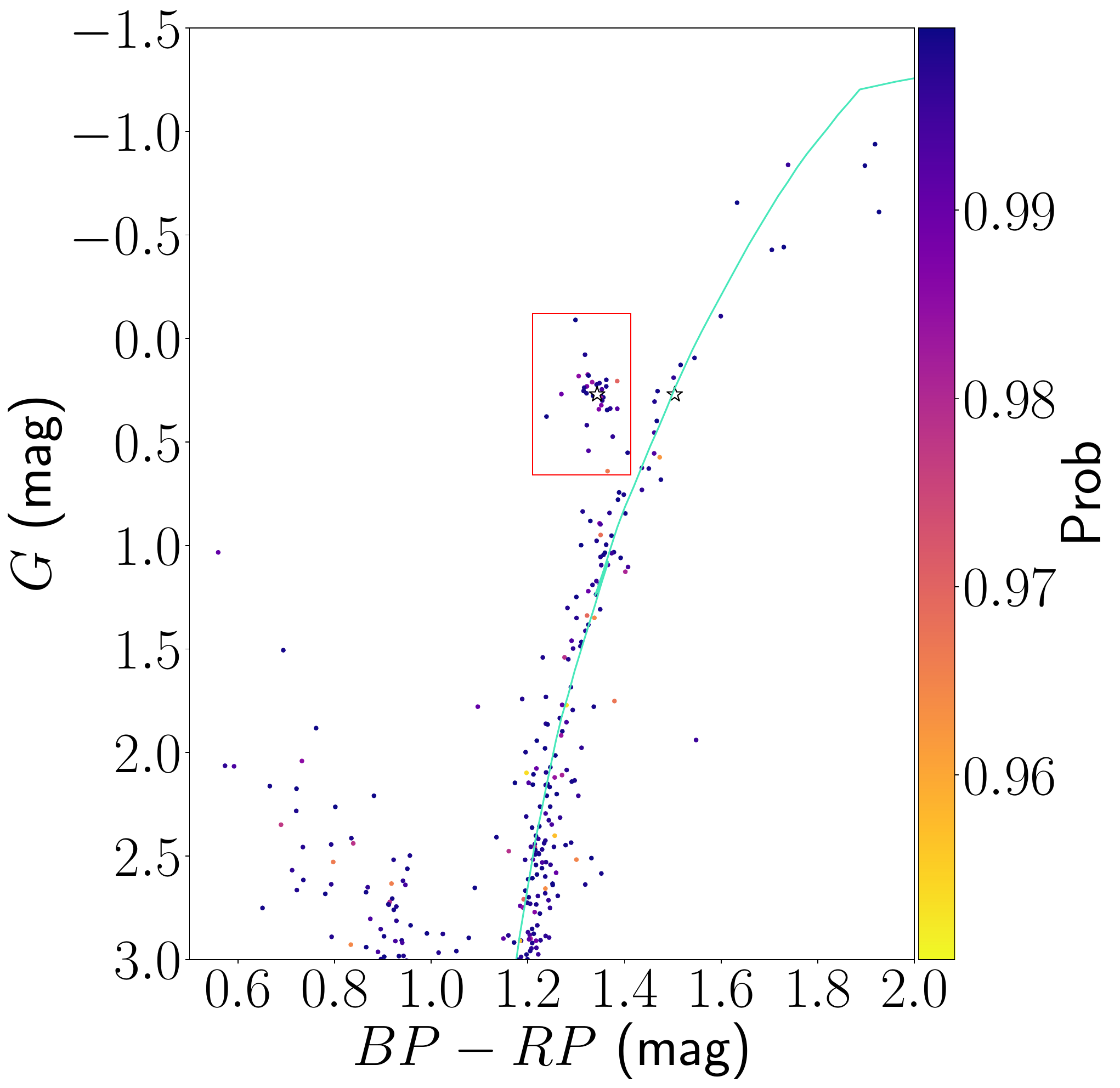}{0.31\textwidth}{(l) NGC~6791.}
    }
    \gridline{ \fig{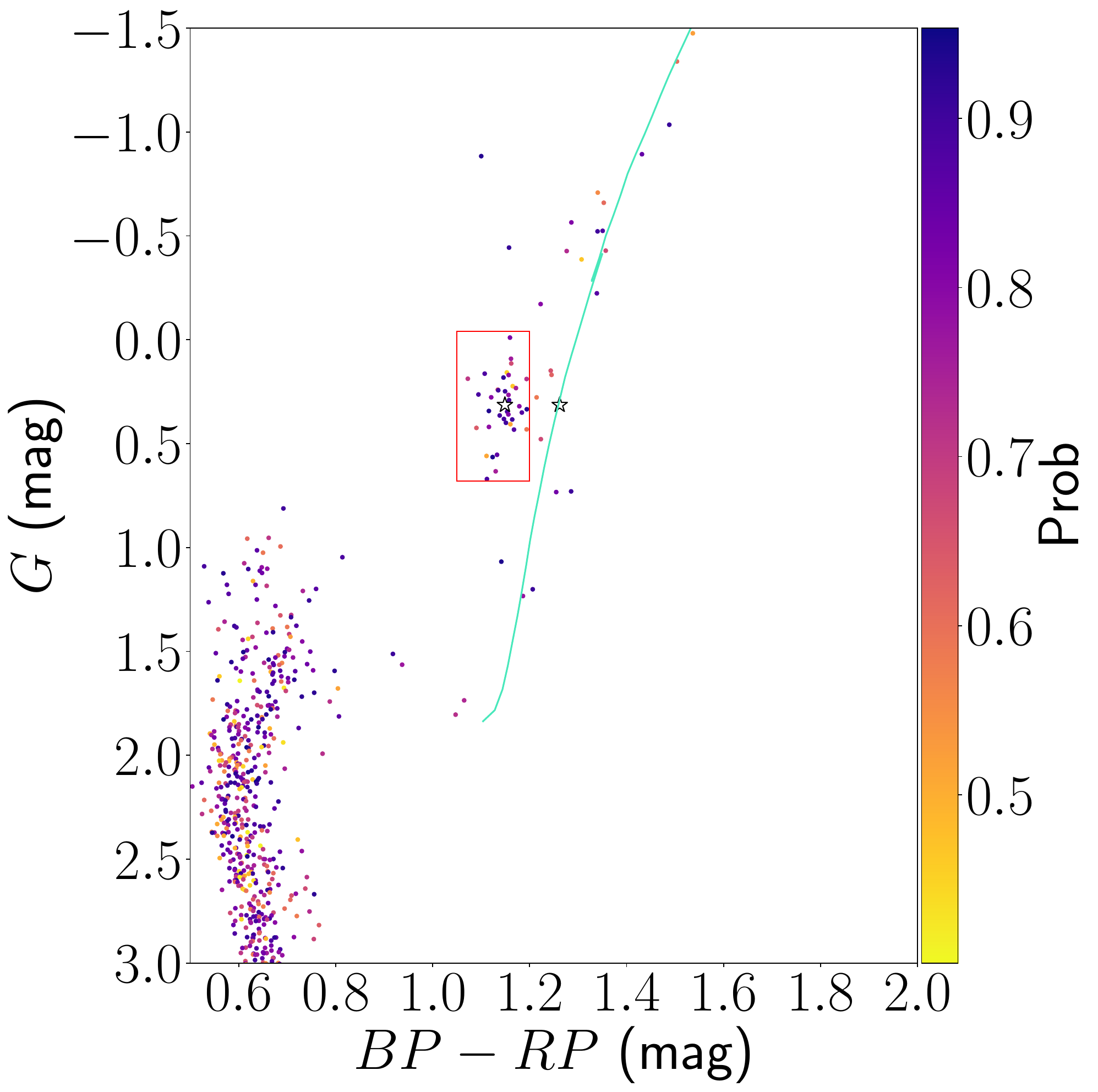}{0.3\textwidth}{(m) NGC~7789.}
      \fig{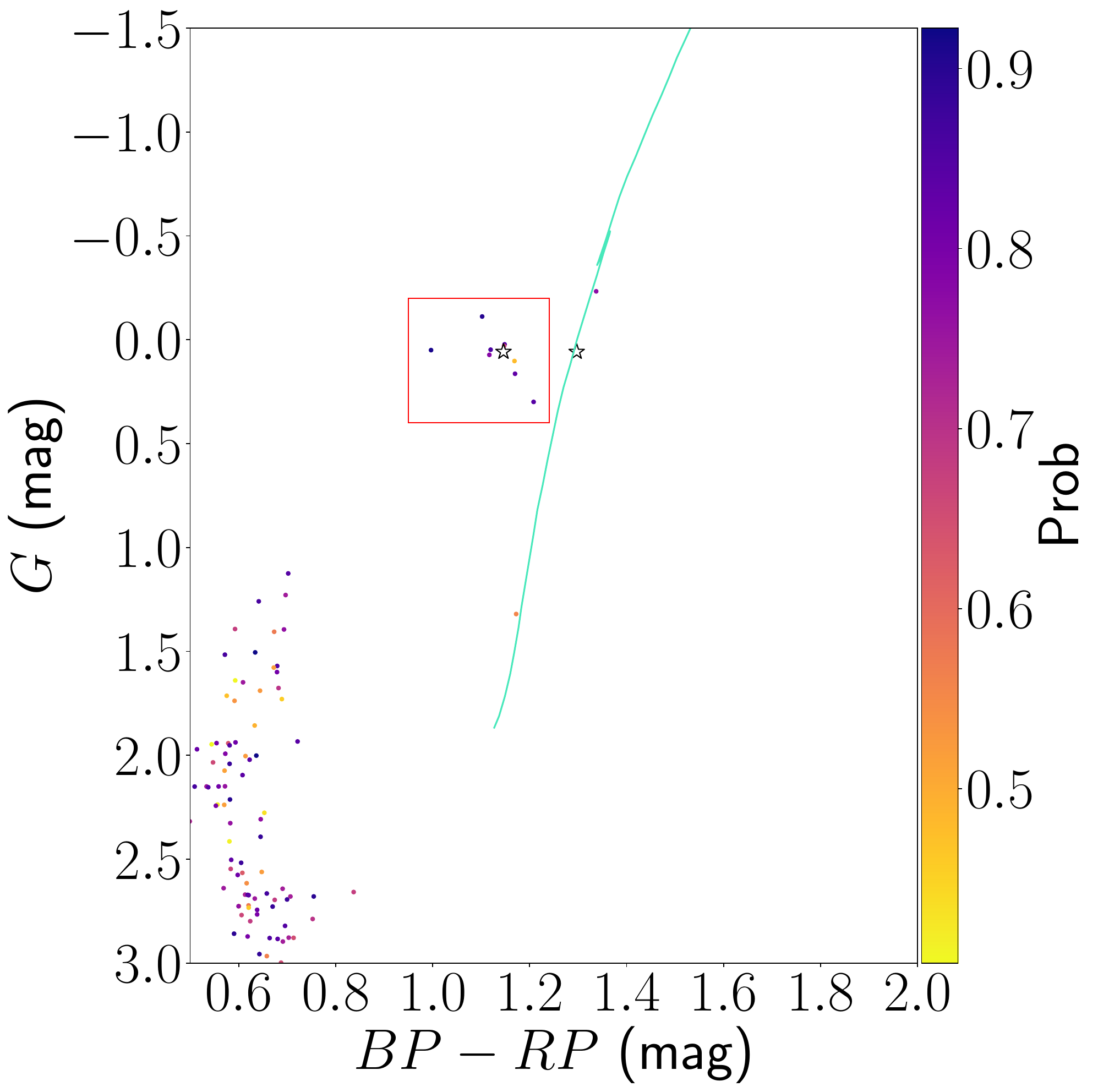}{0.3\textwidth}{(n) Ruprecht~68.}
    }
    \caption{CMDs of galactic open clusters continued from Figure \ref{fig:rc_rgb_cmd_full}.} 
    \label{fig:rc_rgb_cmd_full_2}
\end{figure}

\begin{figure}[hp]
    \gridline{ \fig{figures/SMC26_cmd.pdf}{0.3\textwidth}{(a) Field around SMC~26.}
      \fig{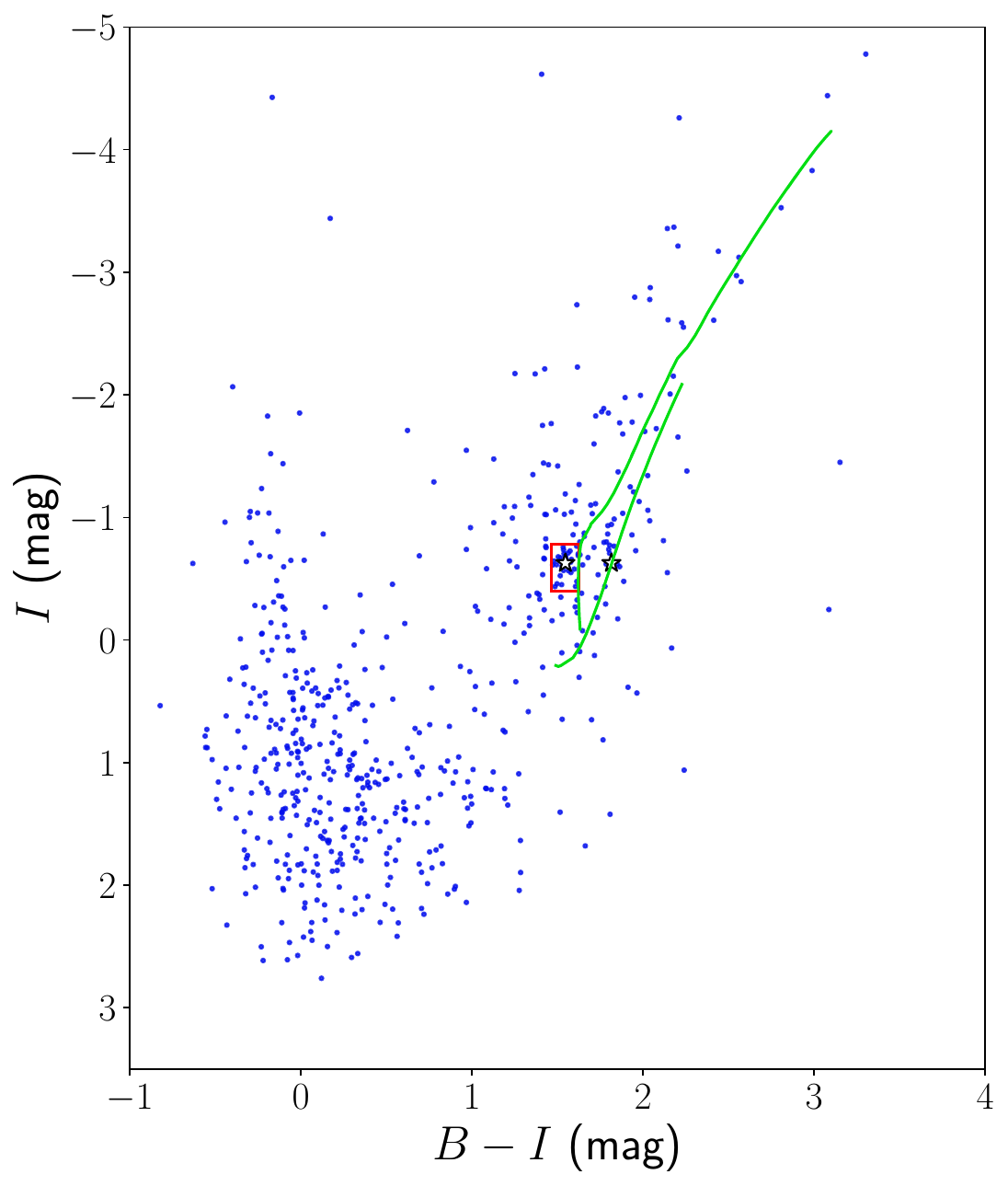}{0.3\textwidth}{(b) Field around SMC~225.}
      \fig{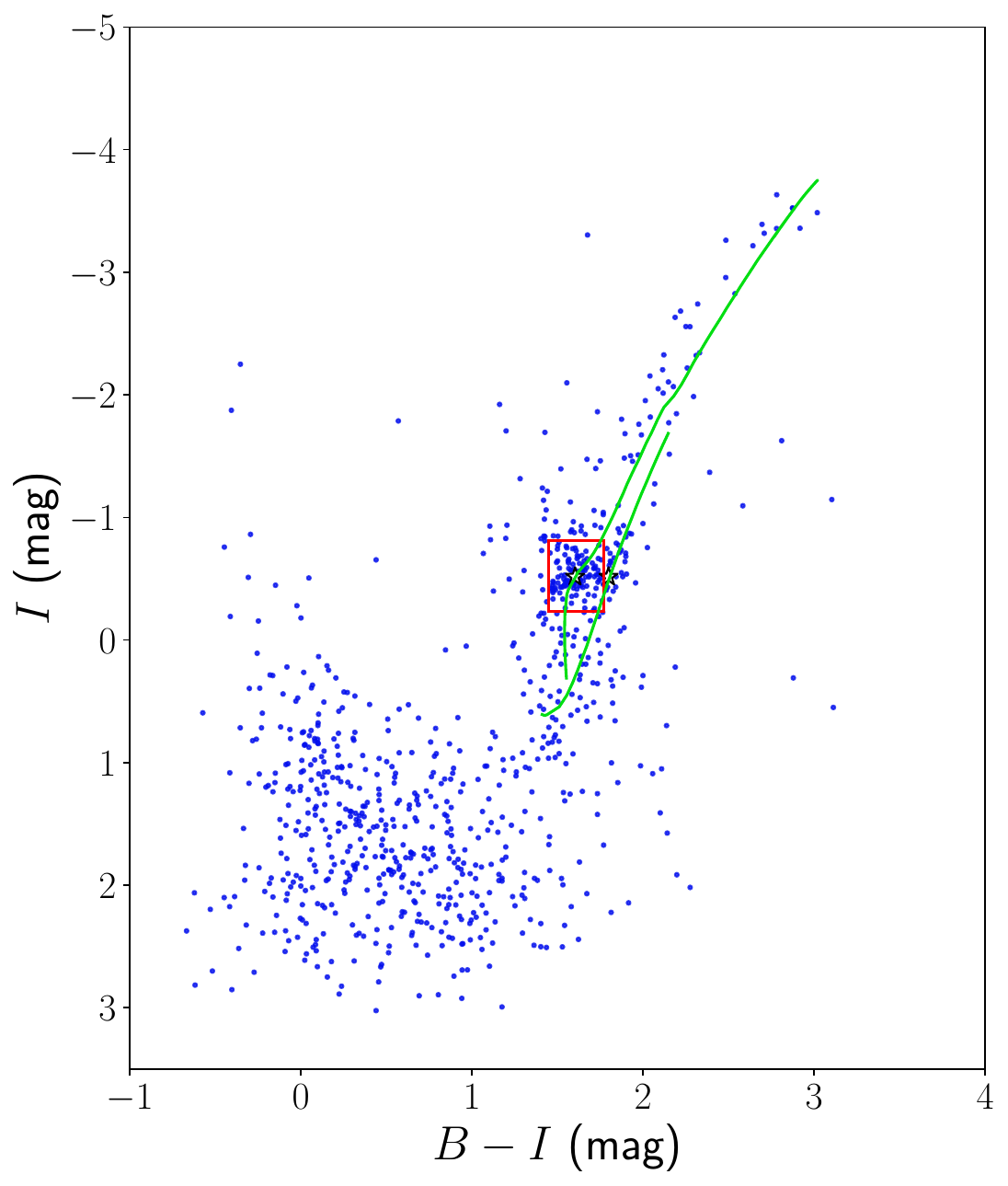}{0.3\textwidth}{(c) Field around SMC~245.}
    }
    \gridline{ \fig{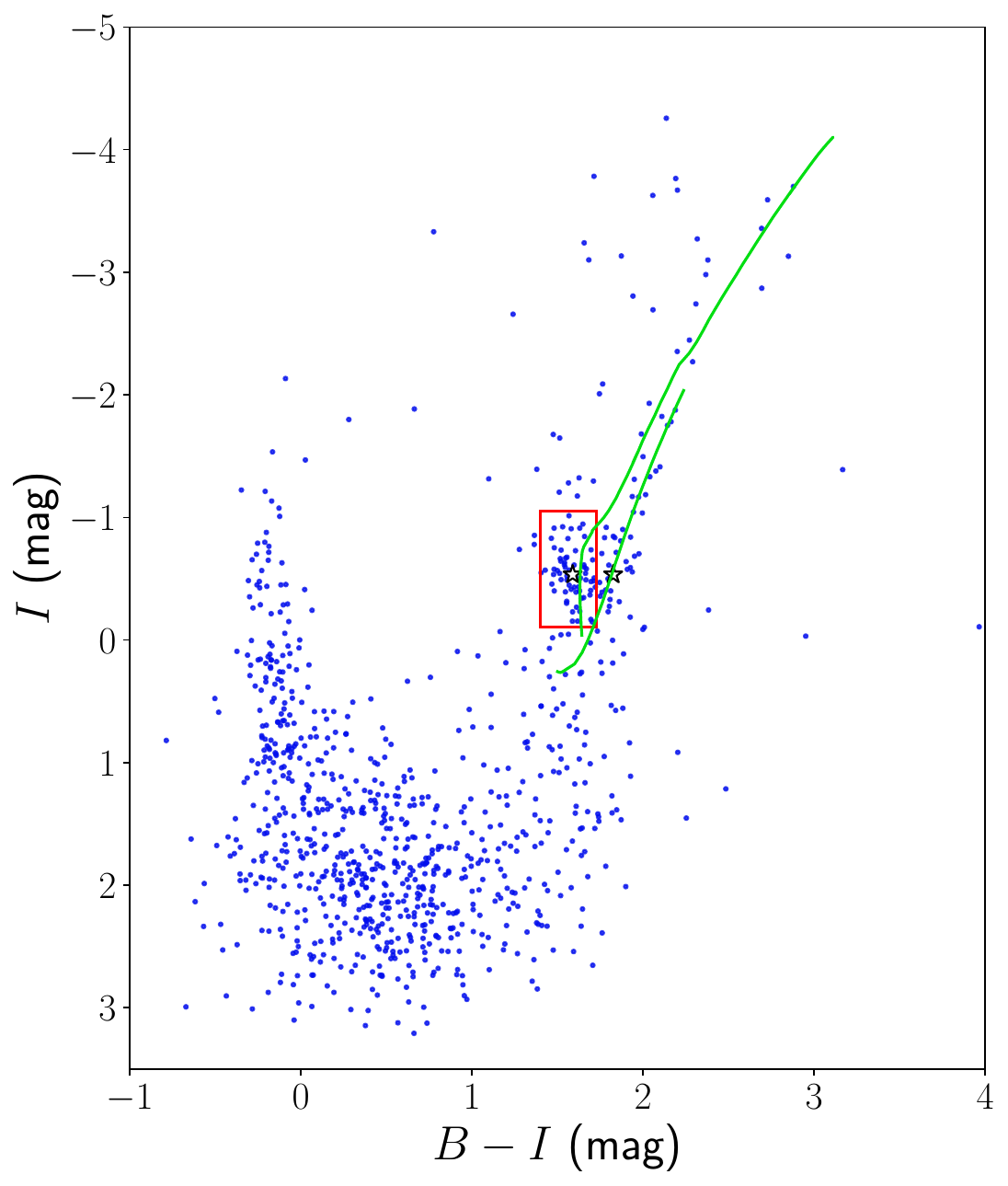}{0.3\textwidth}{(d) Field around SMC~368.}
      \fig{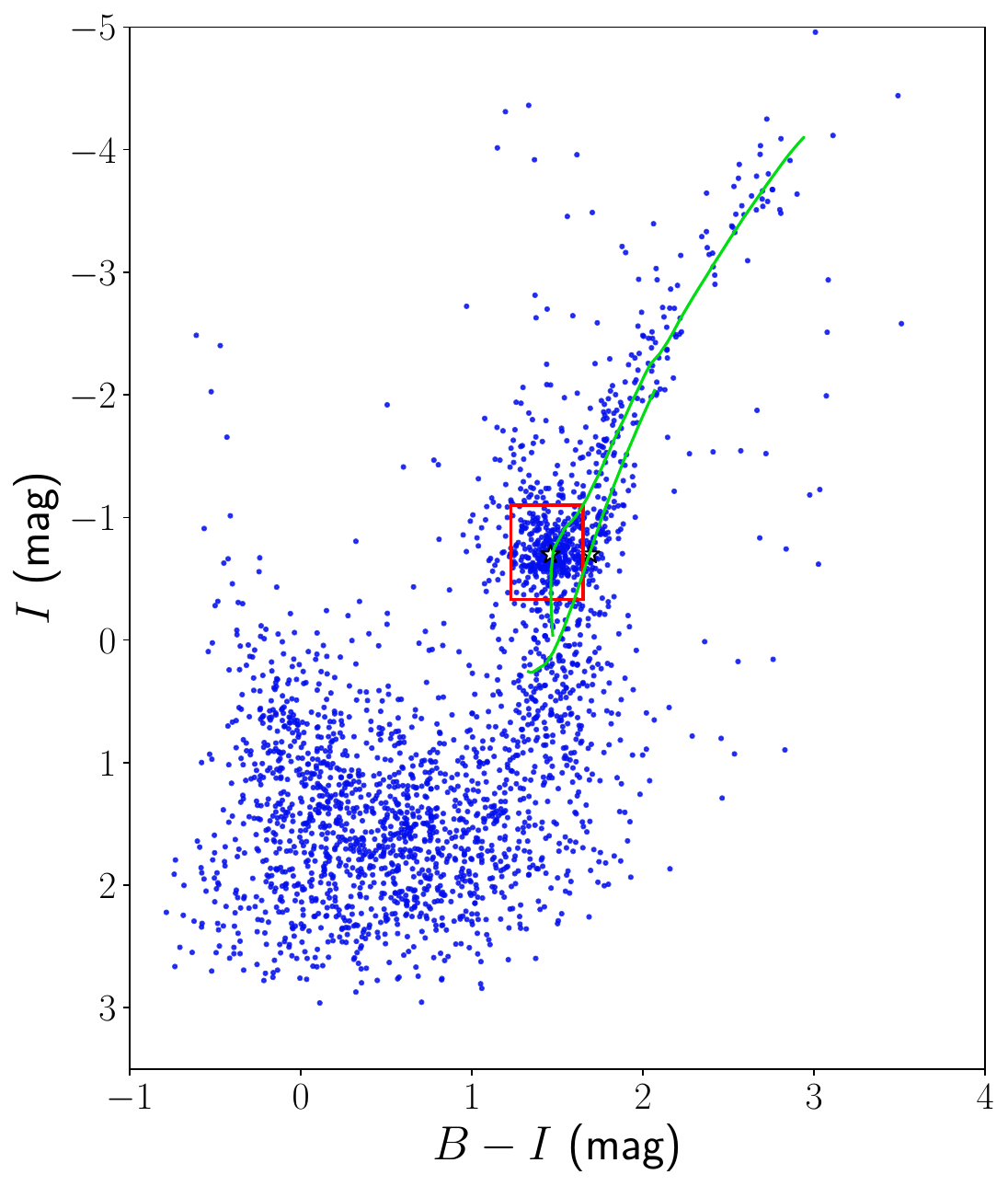}{0.3\textwidth}{(e) Field around SMC~571.}
    }
    \caption{CMDs of fields surrounding clusters in the SMC. Symbols and colors are the same as in Figure \ref{fig:smc_cmd}.} 
    \label{fig:smc_cmd_full_2}
\end{figure}

The CMDs of all fourteen Milky Way open clusters and five SMC fields used in this work are presented in Figs \ref{fig:rc_rgb_cmd_full}, \ref{fig:rc_rgb_cmd_full_2}, and \ref{fig:smc_cmd_full_2}. In each figure, we zoom in on the RGB and include the estimated boundary of the red clump, the fitted RGB segment of a solar-metallicity PARSEC isochrone, and the red clump and RGB loci to illustrate the method.

%CLUMP/RGB CMD FIGURE

%% For this sample we use BibTeX plus aasjournals.bst to generate the
%% the bibliography. The sample631.bib file was populated from ADS. To
%% get the citations to show in the compiled file do the following:
%%
%% pdflatex sample631.tex
%% bibtext sample631
%% pdflatex sample631.tex
%% pdflatex sample631.tex

\bibliography{clump}{}

\begin{thebibliography}{}
\expandafter\ifx\csname natexlab\endcsname\relax\def\natexlab#1{#1}\fi
\providecommand{\url}[1]{\href{#1}{#1}}
\providecommand{\dodoi}[1]{doi:~\href{http://doi.org/#1}{\nolinkurl{#1}}}
\providecommand{\doeprint}[1]{\href{http://ascl.net/#1}{\nolinkurl{http://ascl.net/#1}}}
\providecommand{\doarXiv}[1]{\href{https://arxiv.org/abs/#1}{\nolinkurl{https://arxiv.org/abs/#1}}}

\bibitem[{{An} {et~al.}(2019){An}, {Pinsonneault}, {Terndrup}, \& {Chung}}]{2019ApJ...879...81A}
{An}, D., {Pinsonneault}, M.~H., {Terndrup}, D.~M., \& {Chung}, C. 2019, \apj, 879, 81, \dodoi{10.3847/1538-4357/ab23ed}

\bibitem[{{Astropy Collaboration} {et~al.}(2013){Astropy Collaboration}, {Robitaille}, {Tollerud}, {Greenfield}, {Droettboom}, {Bray}, {Aldcroft}, {Davis}, {Ginsburg}, {Price-Whelan}, {Kerzendorf}, {Conley}, {Crighton}, {Barbary}, {Muna}, {Ferguson}, {Grollier}, {Parikh}, {Nair}, {Unther}, {Deil}, {Woillez}, {Conseil}, {Kramer}, {Turner}, {Singer}, {Fox}, {Weaver}, {Zabalza}, {Edwards}, {Azalee Bostroem}, {Burke}, {Casey}, {Crawford}, {Dencheva}, {Ely}, {Jenness}, {Labrie}, {Lim}, {Pierfederici}, {Pontzen}, {Ptak}, {Refsdal}, {Servillat}, \& {Streicher}}]{astropy:2013}
{Astropy Collaboration}, {Robitaille}, T.~P., {Tollerud}, E.~J., {et~al.} 2013, \apj, 558, A33, \dodoi{10.1051/0004-6361/201322068}

\bibitem[{{Astropy Collaboration} {et~al.}(2018){Astropy Collaboration}, {Price-Whelan}, {Sip{\H{o}}cz}, {G{\"u}nther}, {Lim}, {Crawford}, {Conseil}, {Shupe}, {Craig}, {Dencheva}, {Ginsburg}, {Vand erPlas}, {Bradley}, {P{\'e}rez-Su{\'a}rez}, {de Val-Borro}, {Aldcroft}, {Cruz}, {Robitaille}, {Tollerud}, {Ardelean}, {Babej}, {Bach}, {Bachetti}, {Bakanov}, {Bamford}, {Barentsen}, {Barmby}, {Baumbach}, {Berry}, {Biscani}, {Boquien}, {Bostroem}, {Bouma}, {Brammer}, {Bray}, {Breytenbach}, {Buddelmeijer}, {Burke}, {Calderone}, {Cano Rodr{\'\i}guez}, {Cara}, {Cardoso}, {Cheedella}, {Copin}, {Corrales}, {Crichton}, {D'Avella}, {Deil}, {Depagne}, {Dietrich}, {Donath}, {Droettboom}, {Earl}, {Erben}, {Fabbro}, {Ferreira}, {Finethy}, {Fox}, {Garrison}, {Gibbons}, {Goldstein}, {Gommers}, {Greco}, {Greenfield}, {Groener}, {Grollier}, {Hagen}, {Hirst}, {Homeier}, {Horton}, {Hosseinzadeh}, {Hu}, {Hunkeler}, {Ivezi{\'c}}, {Jain}, {Jenness}, {Kanarek}, {Kendrew}, {Kern}, {Kerzendorf}, {Khvalko}, {King}, {Kirkby}, {Kulkarni},
  {Kumar}, {Lee}, {Lenz}, {Littlefair}, {Ma}, {Macleod}, {Mastropietro}, {McCully}, {Montagnac}, {Morris}, {Mueller}, {Mumford}, {Muna}, {Murphy}, {Nelson}, {Nguyen}, {Ninan}, {N{\"o}the}, {Ogaz}, {Oh}, {Parejko}, {Parley}, {Pascual}, {Patil}, {Patil}, {Plunkett}, {Prochaska}, {Rastogi}, {Reddy Janga}, {Sabater}, {Sakurikar}, {Seifert}, {Sherbert}, {Sherwood-Taylor}, {Shih}, {Sick}, {Silbiger}, {Singanamalla}, {Singer}, {Sladen}, {Sooley}, {Sornarajah}, {Streicher}, {Teuben}, {Thomas}, {Tremblay}, {Turner}, {Terr{\'o}n}, {van Kerkwijk}, {de la Vega}, {Watkins}, {Weaver}, {Whitmore}, {Woillez}, {Zabalza}, \& {Astropy Contributors}}]{astropy:2018}
{Astropy Collaboration}, {Price-Whelan}, A.~M., {Sip{\H{o}}cz}, B.~M., {et~al.} 2018, \aj, 156, 123, \dodoi{10.3847/1538-3881/aabc4f}

\bibitem[{{Astropy Collaboration} {et~al.}(2022){Astropy Collaboration}, {Price-Whelan}, {Lim}, {Earl}, {Starkman}, {Bradley}, {Shupe}, {Patil}, {Corrales}, {Brasseur}, {N{"o}the}, {Donath}, {Tollerud}, {Morris}, {Ginsburg}, {Vaher}, {Weaver}, {Tocknell}, {Jamieson}, {van Kerkwijk}, {Robitaille}, {Merry}, {Bachetti}, {G{"u}nther}, {Aldcroft}, {Alvarado-Montes}, {Archibald}, {B{'o}di}, {Bapat}, {Barentsen}, {Baz{'a}n}, {Biswas}, {Boquien}, {Burke}, {Cara}, {Cara}, {Conroy}, {Conseil}, {Craig}, {Cross}, {Cruz}, {D'Eugenio}, {Dencheva}, {Devillepoix}, {Dietrich}, {Eigenbrot}, {Erben}, {Ferreira}, {Foreman-Mackey}, {Fox}, {Freij}, {Garg}, {Geda}, {Glattly}, {Gondhalekar}, {Gordon}, {Grant}, {Greenfield}, {Groener}, {Guest}, {Gurovich}, {Handberg}, {Hart}, {Hatfield-Dodds}, {Homeier}, {Hosseinzadeh}, {Jenness}, {Jones}, {Joseph}, {Kalmbach}, {Karamehmetoglu}, {Ka{l}uszy{'n}ski}, {Kelley}, {Kern}, {Kerzendorf}, {Koch}, {Kulumani}, {Lee}, {Ly}, {Ma}, {MacBride}, {Maljaars}, {Muna}, {Murphy}, {Norman}, {O'Steen},
  {Oman}, {Pacifici}, {Pascual}, {Pascual-Granado}, {Patil}, {Perren}, {Pickering}, {Rastogi}, {Roulston}, {Ryan}, {Rykoff}, {Sabater}, {Sakurikar}, {Salgado}, {Sanghi}, {Saunders}, {Savchenko}, {Schwardt}, {Seifert-Eckert}, {Shih}, {Jain}, {Shukla}, {Sick}, {Simpson}, {Singanamalla}, {Singer}, {Singhal}, {Sinha}, {Sip{H{o}}cz}, {Spitler}, {Stansby}, {Streicher}, {{{S}}umak}, {Swinbank}, {Taranu}, {Tewary}, {Tremblay}, {Val-Borro}, {Van Kooten}, {Vasovi{'c}}, {Verma}, {de Miranda Cardoso}, {Williams}, {Wilson}, {Winkel}, {Wood-Vasey}, {Xue}, {Yoachim}, {Zhang}, {Zonca}, \& {Astropy Project Contributors}}]{astropy:2022}
{Astropy Collaboration}, {Price-Whelan}, A.~M., {Lim}, P.~L., {et~al.} 2022, \apj, 935, 167, \dodoi{10.3847/1538-4357/ac7c74}

\bibitem[{{Balaguer-N{\'u}nez} {et~al.}(1998){Balaguer-N{\'u}nez}, {Tian}, \& {Zhao}}]{1998A&AS..133..387B}
{Balaguer-N{\'u}nez}, L., {Tian}, K.~P., \& {Zhao}, J.~L. 1998, \aaps, 133, 387, \dodoi{10.1051/aas:1998324}

\bibitem[{{Beccari} {et~al.}(2012){Beccari}, {L{\"u}tzgendorf}, {Olczak}, {Ferraro}, {Lanzoni}, {Carraro}, {Stetson}, {Sollima}, \& {Boffin}}]{2012ApJ...754..108B}
{Beccari}, G., {L{\"u}tzgendorf}, N., {Olczak}, C., {et~al.} 2012, \apj, 754, 108, \dodoi{10.1088/0004-637X/754/2/108}

\bibitem[{{Beers} {et~al.}(1990){Beers}, {Flynn}, \& {Gebhardt}}]{1990AJ....100...32B}
{Beers}, T.~C., {Flynn}, K., \& {Gebhardt}, K. 1990, \aj, 100, 32, \dodoi{10.1086/115487}

\bibitem[{{Bomans} {et~al.}(1995){Bomans}, {Vallenari}, \& {de Boer}}]{1995A&A...298..427B}
{Bomans}, D.~J., {Vallenari}, A., \& {de Boer}, K.~S. 1995, \aap, 298, 427

\bibitem[{{Bressan} {et~al.}(1994){Bressan}, {Chiosi}, \& {Fagotto}}]{1994ApJS...94...63B}
{Bressan}, A., {Chiosi}, C., \& {Fagotto}, F. 1994, \apjs, 94, 63, \dodoi{10.1086/192073}

\bibitem[{{Bressan} {et~al.}(2012){Bressan}, {Marigo}, {Girardi}, {Salasnich}, {Dal Cero}, {Rubele}, \& {Nanni}}]{2012MNRAS.427..127B}
{Bressan}, A., {Marigo}, P., {Girardi}, L., {et~al.} 2012, \mnras, 427, 127, \dodoi{10.1111/j.1365-2966.2012.21948.x}

\bibitem[{{Cantat-Gaudin} {et~al.}(2020){Cantat-Gaudin}, {Anders}, {Castro-Ginard}, {Jordi}, {Romero-G{\'o}mez}, {Soubiran}, {Casamiquela}, {Tarricq}, {Moitinho}, {Vallenari}, {Bragaglia}, {Krone-Martins}, \& {Kounkel}}]{2020A&A...640A...1C}
{Cantat-Gaudin}, T., {Anders}, F., {Castro-Ginard}, A., {et~al.} 2020, \aap, 640, A1, \dodoi{10.1051/0004-6361/202038192}

\bibitem[{{Casagrande} \& {VandenBerg}(2018)}]{2018MNRAS.479L.102C}
{Casagrande}, L., \& {VandenBerg}, D.~A. 2018, \mnras, 479, L102, \dodoi{10.1093/mnrasl/sly104}

\bibitem[{{Chaboyer}(1995)}]{1995ApJ...444L...9C}
{Chaboyer}, B. 1995, \apjl, 444, L9, \dodoi{10.1086/187847}

\bibitem[{{Conroy} {et~al.}(2014){Conroy}, {Graves}, \& {van Dokkum}}]{2014ApJ...780...33C}
{Conroy}, C., {Graves}, G.~J., \& {van Dokkum}, P.~G. 2014, \apj, 780, 33, \dodoi{10.1088/0004-637X/780/1/33}

\bibitem[{{Da Costa} \& {Hatzidimitriou}(1998)}]{1998AJ....115.1934D}
{Da Costa}, G.~S., \& {Hatzidimitriou}, D. 1998, \aj, 115, 1934, \dodoi{10.1086/300340}

\bibitem[{{degl'Innocenti} {et~al.}(2000){degl'Innocenti}, {Castellani}, {Girardi}, {Marconi}, {Prada Moroni}, \& {Weiss}}]{2000ASPC..211..169D}
{degl'Innocenti}, S., {Castellani}, V., {Girardi}, L., {et~al.} 2000, in Astronomical Society of the Pacific Conference Series, Vol. 211, Massive Stellar Clusters, ed. A.~{Lan{\c{c}}on} \& C.~M. {Boily}, 169

\bibitem[{{Deng} {et~al.}(1999){Deng}, {Chen}, {Liu}, \& {Chen}}]{1999ApJ...524..824D}
{Deng}, L., {Chen}, R., {Liu}, X.~S., \& {Chen}, J.~S. 1999, \apj, 524, 824, \dodoi{10.1086/307832}

\bibitem[{{Dias} {et~al.}(2021){Dias}, {Monteiro}, {Moitinho}, {L{\'e}pine}, {Carraro}, {Paunzen}, {Alessi}, \& {Villela}}]{2021MNRAS.504..356D}
{Dias}, W.~S., {Monteiro}, H., {Moitinho}, A., {et~al.} 2021, \mnras, 504, 356, \dodoi{10.1093/mnras/stab770}

\bibitem[{{Dolphin}(2002)}]{2002MNRAS.332...91D}
{Dolphin}, A.~E. 2002, \mnras, 332, 91, \dodoi{10.1046/j.1365-8711.2002.05271.x}

\bibitem[{{Dotter} {et~al.}(2007){Dotter}, {Chaboyer}, {Ferguson}, {Lee}, {Worthey}, {Jevremovi{\'c}}, \& {Baron}}]{2007ApJ...666..403D}
{Dotter}, A., {Chaboyer}, B., {Ferguson}, J.~W., {et~al.} 2007, \apj, 666, 403, \dodoi{10.1086/519946}

\bibitem[{{Girardi}(2016)}]{2016ARA&A..54...95G}
{Girardi}, L. 2016, \araa, 54, 95, \dodoi{10.1146/annurev-astro-081915-023354}

\bibitem[{{Girardi} {et~al.}(2000){Girardi}, {Mermilliod}, \& {Carraro}}]{2000A&A...354..892G}
{Girardi}, L., {Mermilliod}, J.~C., \& {Carraro}, G. 2000, \aap, 354, 892, \dodoi{10.48550/arXiv.astro-ph/0001068}

\bibitem[{{Glatt} {et~al.}(2010){Glatt}, {Grebel}, \& {Koch}}]{2010A&A...517A..50G}
{Glatt}, K., {Grebel}, E.~K., \& {Koch}, A. 2010, \aap, 517, A50, \dodoi{10.1051/0004-6361/201014187}

\bibitem[{{Gratton} {et~al.}(2012){Gratton}, {Carretta}, \& {Bragaglia}}]{2012A&ARv..20...50G}
{Gratton}, R.~G., {Carretta}, E., \& {Bragaglia}, A. 2012, \aapr, 20, 50, \dodoi{10.1007/s00159-012-0050-3}

\bibitem[{{Griggio} \& {Bedin}(2022)}]{2022MNRAS.511.4702G}
{Griggio}, M., \& {Bedin}, L.~R. 2022, \mnras, 511, 4702, \dodoi{10.1093/mnras/stac391}

\bibitem[{{Hansen}(2005)}]{2005ApJ...635..522H}
{Hansen}, B. M.~S. 2005, \apj, 635, 522, \dodoi{10.1086/496951}

\bibitem[{{Harris} \& {Zaritsky}(2001)}]{2001ApJS..136...25H}
{Harris}, J., \& {Zaritsky}, D. 2001, \apjs, 136, 25, \dodoi{10.1086/321792}

\bibitem[{{Harris} \& {Zaritsky}(2004)}]{2004AJ....127.1531H}
---. 2004, \aj, 127, 1531, \dodoi{10.1086/381953}

\bibitem[{{Harris}(1996)}]{1996AJ....112.1487H}
{Harris}, W.~E. 1996, \aj, 112, 1487, \dodoi{10.1086/118116}

\bibitem[{{Hatzidimitriou}(1991)}]{1991MNRAS.251..545H}
{Hatzidimitriou}, D. 1991, \mnras, 251, 545, \dodoi{10.1093/mnras/251.4.545}

\bibitem[{{Hidalgo} {et~al.}(2018){Hidalgo}, {Pietrinferni}, {Cassisi}, {Salaris}, {Mucciarelli}, {Savino}, {Aparicio}, {Silva Aguirre}, \& {Verma}}]{2018ApJ...856..125H}
{Hidalgo}, S.~L., {Pietrinferni}, A., {Cassisi}, S., {et~al.} 2018, \apj, 856, 125, \dodoi{10.3847/1538-4357/aab158}

\bibitem[{{Idiart} {et~al.}(2007){Idiart}, {Maciel}, \& {Costa}}]{2007A&A...472..101I}
{Idiart}, T.~P., {Maciel}, W.~J., \& {Costa}, R.~D.~D. 2007, \aap, 472, 101, \dodoi{10.1051/0004-6361:20077674}

\bibitem[{{Kerber} {et~al.}(2007){Kerber}, {Santiago}, \& {Brocato}}]{2007A&A...462..139K}
{Kerber}, L.~O., {Santiago}, B.~X., \& {Brocato}, E. 2007, \aap, 462, 139, \dodoi{10.1051/0004-6361:20066128}

\bibitem[{{Kharchenko} {et~al.}(2013){Kharchenko}, {Piskunov}, {Schilbach}, {R{\"o}ser}, \& {Scholz}}]{2013A&A...558A..53K}
{Kharchenko}, N.~V., {Piskunov}, A.~E., {Schilbach}, E., {R{\"o}ser}, S., \& {Scholz}, R.~D. 2013, \aap, 558, A53, \dodoi{10.1051/0004-6361/201322302}

\bibitem[{{Marigo} {et~al.}(2008){Marigo}, {Girardi}, {Bressan}, {Groenewegen}, {Silva}, \& {Granato}}]{2008A&A...482..883M}
{Marigo}, P., {Girardi}, L., {Bressan}, A., {et~al.} 2008, \aap, 482, 883, \dodoi{10.1051/0004-6361:20078467}

\bibitem[{{Mighell} {et~al.}(1998){Mighell}, {Sarajedini}, \& {French}}]{1998AJ....116.2395M}
{Mighell}, K.~J., {Sarajedini}, A., \& {French}, R.~S. 1998, \aj, 116, 2395, \dodoi{10.1086/300591}

\bibitem[{{Olszewski} {et~al.}(1991){Olszewski}, {Schommer}, {Suntzeff}, \& {Harris}}]{1991AJ....101..515O}
{Olszewski}, E.~W., {Schommer}, R.~A., {Suntzeff}, N.~B., \& {Harris}, H.~C. 1991, \aj, 101, 515, \dodoi{10.1086/115701}

\bibitem[{{Pastorelli} {et~al.}(2020){Pastorelli}, {Marigo}, {Girardi}, {Aringer}, {Chen}, {Rubele}, {Trabucchi}, {Bladh}, {Boyer}, {Bressan}, {Dalcanton}, {Groenewegen}, {Lebzelter}, {Mowlavi}, {Chubb}, {Cioni}, {de Grijs}, {Ivanov}, {Nanni}, {van Loon}, \& {Zaggia}}]{2020MNRAS.498.3283P}
{Pastorelli}, G., {Marigo}, P., {Girardi}, L., {et~al.} 2020, \mnras, 498, 3283, \dodoi{10.1093/mnras/staa2565}

\bibitem[{Pieres {et~al.}(2016)Pieres, Santiago, Balbinot, Luque, Queiroz, da Costa, Maia, Drlica-Wagner, Roodman, Abbott, Allam, Benoit-Lévy, Bertin, Brooks, Buckley-Geer, Burke, Rosell, Kind, Carretero, Cunha, Desai, Diehl, Eifler, Finley, Flaugher, Fosalba, Frieman, Gerdes, Gruen, Gruendl, Gutierrez, Honscheid, James, Kuehn, Kuropatkin, Lahav, Li, Marshall, Martini, Miller, Miquel, Nichol, Nord, Ogando, Plazas, Romer, Sanchez, Scarpine, Schubnell, Sevilla-Noarbe, Smith, Soares-Santos, Sobreira, Suchyta, Swanson, Tarle, Thaler, Thomas, Tucker, \& Walker}]{10.1093/mnras/stw1260}
Pieres, A., Santiago, B., Balbinot, E., {et~al.} 2016, Monthly Notices of the Royal Astronomical Society, 461, 519, \dodoi{10.1093/mnras/stw1260}

\bibitem[{{Pietrinferni} {et~al.}(2004){Pietrinferni}, {Cassisi}, {Salaris}, \& {Castelli}}]{2004ApJ...612..168P}
{Pietrinferni}, A., {Cassisi}, S., {Salaris}, M., \& {Castelli}, F. 2004, \apj, 612, 168, \dodoi{10.1086/422498}

\bibitem[{{Reimers}(1975)}]{1975MSRSL...8..369R}
{Reimers}, D. 1975, Memoires of the Societe Royale des Sciences de Liege, 8, 369

\bibitem[{{Riello} {et~al.}(2021){Riello}, {De Angeli}, {Evans}, {Montegriffo}, {Carrasco}, {Busso}, {Palaversa}, {Burgess}, {Diener}, {Davidson}, {Rowell}, {Fabricius}, {Jordi}, {Bellazzini}, {Pancino}, {Harrison}, {Cacciari}, {van Leeuwen}, {Hambly}, {Hodgkin}, {Osborne}, {Altavilla}, {Barstow}, {Brown}, {Castellani}, {Cowell}, {De Luise}, {Gilmore}, {Giuffrida}, {Hidalgo}, {Holland}, {Marinoni}, {Pagani}, {Piersimoni}, {Pulone}, {Ragaini}, {Rainer}, {Richards}, {Sanna}, {Walton}, {Weiler}, \& {Yoldas}}]{2021A&A...649A...3R}
{Riello}, M., {De Angeli}, F., {Evans}, D.~W., {et~al.} 2021, \aap, 649, A3, \dodoi{10.1051/0004-6361/202039587}

\bibitem[{Virtanen {et~al.}(2020)Virtanen, Gommers, Oliphant, Haberland, Reddy, Cournapeau, Burovski, Peterson, Weckesser, Bright, {van der Walt}, Brett, Wilson, Millman, Mayorov, Nelson, Jones, Kern, Larson, Carey, Polat, Feng, Moore, {VanderPlas}, Laxalde, Perktold, Cimrman, Henriksen, Quintero, Harris, Archibald, Ribeiro, Pedregosa, {van Mulbregt}, \& {SciPy 1.0 Contributors}}]{2020SciPy-NMeth}
Virtanen, P., Gommers, R., Oliphant, T.~E., {et~al.} 2020, Nature Methods, 17, 261, \dodoi{10.1038/s41592-019-0686-2}

\bibitem[{{Worthey}(1994)}]{1994ApJS...95..107W}
{Worthey}, G. 1994, \apjs, 95, 107, \dodoi{10.1086/192096}

\bibitem[{{Worthey} \& {Lee}(2011)}]{2011ApJS..193....1W}
{Worthey}, G., \& {Lee}, H.-c. 2011, \apjs, 193, 1, \dodoi{10.1088/0067-0049/193/1/1}

\bibitem[{{Worthey} \& {Shi}(2023)}]{2023MNRAS.518.4106W}
{Worthey}, G., \& {Shi}, X. 2023, \mnras, 518, 4106, \dodoi{10.1093/mnras/stac3297}

\bibitem[{{Worthey} {et~al.}(2022){Worthey}, {Shi}, {Pal}, {Lee}, \& {Tang}}]{2022MNRAS.511.3198W}
{Worthey}, G., {Shi}, X., {Pal}, T., {Lee}, H.-c., \& {Tang}, B. 2022, \mnras, 511, 3198, \dodoi{10.1093/mnras/stac267}

\bibitem[{{Zaritsky} {et~al.}(2002){Zaritsky}, {Harris}, {Thompson}, {Grebel}, \& {Massey}}]{2002AJ....123..855Z}
{Zaritsky}, D., {Harris}, J., {Thompson}, I.~B., {Grebel}, E.~K., \& {Massey}, P. 2002, \aj, 123, 855, \dodoi{10.1086/338437}

\end{thebibliography}
\bibliographystyle{aasjournal}

%% This command is needed to show the entire author+affiliation list when
%% the collaboration and author truncation commands are used.  It has to
%% go at the end of the manuscript.
%\allauthors

%% Include this line if you are using the \added, \replaced, \deleted
%% commands to see a summary list of all changes at the end of the article.
%\listofchanges

\end{document}